\documentclass[lettersize,journal]{IEEEtran}
\usepackage{array}
\usepackage{stfloats}
\usepackage{url}
\usepackage{verbatim}
\usepackage{graphicx}
\usepackage{cite}

\usepackage{mathtools}
\usepackage{algorithm}
\usepackage[noend]{algpseudocode}
\usepackage{subfigure}
\usepackage{makecell}
\usepackage{multirow}
\usepackage{amsmath}
\usepackage{amsthm}
\usepackage{amsfonts}
\usepackage{dsfont}
\usepackage{pifont}
\usepackage{hyperref}
\usepackage{fontawesome5} 
\usepackage{textcomp}
\usepackage{enumitem}

\newtheorem{definition}{\bf Definition}

\newtheorem{example}{\bf Example}

\newtheorem{strategy}{\bf Strategy}
\newtheorem{lemma}{\bf Lemma}
\newtheorem*{research}{Research Problem}

\usepackage{ifsym}

\begin{document}
\title{\texttt{ACTIVE}: Continuous Similarity Search for Vessel Trajectories}

\author{Tiantian Liu$^\dagger$, Hengyu Liu$^\dagger$, Tianyi Li$^{\ddagger}$, Kristian Torp, Christian S. Jensen
\thanks{The authors are with Department of Computer Science, Aalborg University, Denmark. (E-mails: \{liutt, heli, tianyi, torp, csj\}@cs.aau.dk)\\
$^\dagger$ Tiantian Liu and Hengyu Liu have equal contributions.\\
$^{\ddagger}$ Tianyi Li is the corresponding author.}}

\markboth{IEEE Transactions on Knowledge and Data Engineering,~Vol.~XX, No.~XX, XXXX}%
{Liu \MakeLowercase{\textit{et al.}}: \texttt{ACTIVE}: Continuous Similarity Search for Vessel Trajectories}

\IEEEpubid{0000--0000/00\$00.00~\copyright~2025 IEEE}
\maketitle

\begin{abstract}
Publicly available vessel trajectory data is emitted continuously from the global AIS system. Continuous trajectory similarity search on this data has applications in, e.g., maritime navigation and safety.
Existing proposals typically assume an offline setting and focus on finding similarities between complete trajectories. Such proposals are less effective when applied to online scenarios, where similarity comparisons must be performed continuously as new trajectory data arrives and trajectories evolve.
We therefore propose a re\underline{\textbf{a}}l-time \underline{\textbf{c}}ontinuous \underline{\textbf{t}}rajectory s\underline{\textbf{i}}milarity search method for \underline{\textbf{ve}}ssels  (\texttt{ACTIVE}). We introduce a novel similarity measure, object-trajectory real-time distance, that emphasizes the anticipated future movement trends of vessels, enabling more predictive and forward-looking comparisons. 
Next, we propose an efficient continuous similar trajectory search (\texttt{CSTS}) algorithm together with a segment-based vessel trajectory index and a variety of search space pruning strategies that reduce unnecessary computations during the continuous similarity search, thereby further improving efficiency. 
Extensive experiments on two large real-world AIS datasets offer evidence that \texttt{ACTIVE} is capable of outperforming state-of-the-art methods considerably. \texttt{ACTIVE} significantly reduces index construction costs and index size while achieving a 70\% reduction in terms of query time and a 60\% increase in terms of hit rate.
\end{abstract}

\section{Introduction}
\label{sec:intro}

\IEEEPARstart{A}{utomatic}
Identification System (AIS) data, which provides real-time information on vessel positions, speeds, and directions, offers a rich resource for analyzing vessel movements. With the growing availability of AIS data, understanding vessel trajectories has become essential for navigation safety~\cite{li2024vessel} and environmental monitoring~\cite{jin2021overview}. Trajectory similarity search~\cite{hu2023spatio} plays a key role in applications such as collision avoidance~\cite{li2022evolutionary}, route optimization~\cite{xu2024managing}, and movement pattern analysis~\cite{hu2024estimator}.

\begin{figure}[!htbp]
    \centering
    \includegraphics[width=0.9\columnwidth]{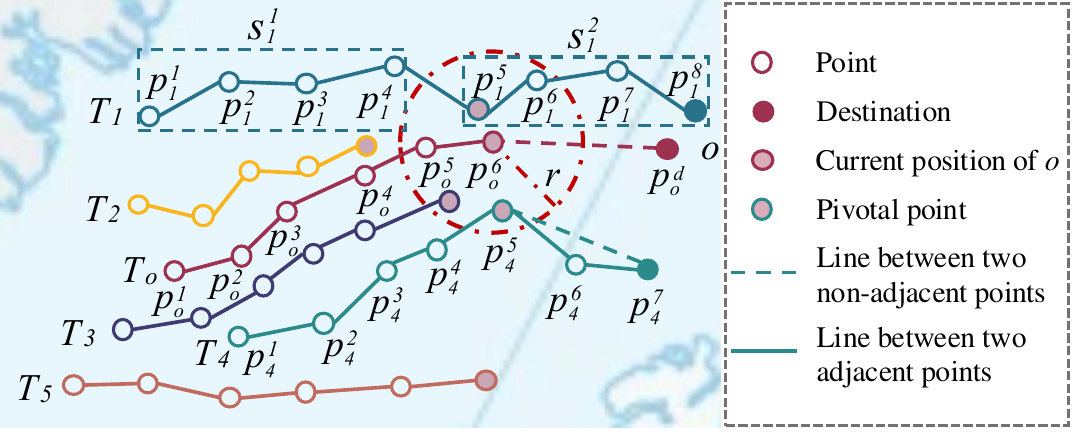}
    \caption{Example of trajectories at timestamp $t$.}
    \label{fig:trajectory}
\end{figure}

Given the dynamic nature of maritime environments, real-time continuous trajectory similarity search is particularly valuable. It enables timely and proactive decision-making by identifying similar movement patterns as new AIS data arrives, facilitating responsive navigation and operational safety. Unlike traditional methods, which rely on retrospective analysis of complete trajectories, real-time similarity search operates incrementally, processing incoming data without requiring an entire trajectory. Moreover, real-time search considers the future movement trends and the destination of a moving object.
For instance, as illustrated in Fig.~\ref{fig:trajectory}, for a moving vessel $o$, given its current trajectory $T_o$ and target destination $p_o^d$, we continuously search for similar historical trajectories, accounting for both past movements and the vessel’s future trend. 
Despite significant progress in real-time trajectory similarity search, existing studies~\cite{ding2008efficient, xie2017distributed, wang2018torch, li2020vessel, he2022trass, luo2023vessel, bakalov2005efficient, bakalov2005time, ta2017signature, shang2017trajectory} face limitations, leading to the following challenges. 

\noindent \textit{\textbf{Challenge I: How to effectively measure trajectory similarity.}} 
Existing methods, such as Dynamic Time Warping~\cite{yi1998efficient, ying2016simple}, Longest Common Subsequence~\cite{vlachos2002discovering}, Fr{\'e}chet distance \cite{alt1995computing, de2013fast, agarwal2014computing}, Hausdorff distance \cite{hangouet1995computation, bai2011polyline}, and 
Edit distance-based measures \cite{chen2004marriage, chen2005robust, ranu2015indexing}, are commonly used to identify similar trajectories by analyzing historical movement data. 
However, they rely heavily on static or full-path comparisons, making them of low utility for evolving trajectories that need to be analyzed incrementally as new data arrives. 
Next, they focus exclusively on historical movements, overlooking the importance of an object's near-future movement and target destination. 
This limits their effectiveness in real-time applications, such as collision detection, traffic management, and route prediction, where anticipating future movements is crucial.
For example, in Fig.~\ref{fig:trajectory}, the most similar trajectory for $T_o$ is $T_3$ using Fr{\'e}chet or Hausdorff distance. However, when considering the destination of the object, the most similar trajectory should be $T_1$. 
\smallskip

\IEEEpubidadjcol
\noindent \textit{\textbf{Challenge II: How to improve efficiency and scalability in continuous similarity trajectory search.}}
Existing similar trajectory search methods~\cite{ding2008efficient, xie2017distributed, wang2018torch, li2020vessel, he2022trass, luo2023vessel} are usually designed for searches using traditional similarity metrics intended for historical data, making them less effective for real-time applications. 
Moreover, they are designed for offline, batch processing on static datasets, requiring full re-computations when new data arrives. This is unsuitable in a setting with continuous updates to AIS data. 
Achieving efficient, real-time continuous similarity searches for vessel trajectories involves addressing several interconnected challenges. 
First, vessel trajectories are complex, often involving irregular and dynamic movements in open water. In contrast, on road networks, structured paths and intersections simplify trajectory analysis.
Second, pruning irrelevant or low-similarity candidates with high accuracy is particularly complex in a dynamic environment where trajectory data changes continuously. 
Effective pruning methods must strike a balance between reducing candidate sets and maintaining the accuracy of similarity computations.
Third, scaling a system and retaining fast response times when handling continuous queries over extended periods adds another layer of difficulty.
The system must be capable of managing large-scale, real-time data streams while maintaining high accuracy and adapting to varying query loads. 
Together, these challenges call for new innovations to achieve a scalable and accurate trajectory search system.

Thus, we present \texttt{ACTIVE}, a novel framework for real-time continuous trajectory similarity search for vessels.

\noindent \textbf{\textit{To address Challenge I}}, we propose a novel similarity measure called object-trajectory real-time distance, which prioritizes the future trends of vessel movements and their target destinations while accounting also for past trajectory parts and the significance of individual trajectory points. This measure enables more accurate predictions of future vessel positions and behaviors, making it better suited for real-time decision-making in maritime navigation and traffic management.

\noindent \textbf{\textit{To address Challenge II}}, we 
design an efficient continuous similar trajectory search algorithm on a segment-based trajectory index with four speed-up strategies that exploit geometric properties, utilize intermediate data, such as similarities computed at previous timestamps, and adjust trajectory granularity. These strategies efficiently eliminate irrelevant trajectory comparisons early in the search process while maintaining accuracy. Together, they optimize continuous searches and reduce computational costs notably. The contributions are summarized as follows.
\setlength{\parsep}{0pt}
\begin{itemize}[itemsep=1pt, leftmargin=10pt]
    \item We propose a framework for re\underline{\textbf{a}}l-time \underline{\textbf{c}}ontinuous \underline{\textbf{t}}rajectory s\underline{\textbf{i}}milarity search for \underline{\textbf{ve}}ssels (\texttt{ACTIVE}). To the best of our knowledge, it is the first to support real-time similar trajectory search while considering the future trends of moving vessels.


    \item We introduce a novel similarity measure, the  \underline{\textbf{o}}bject-\underline{\textbf{t}}rajectory \underline{\textbf{r}}eal-time \underline{\textbf{d}}istance (\texttt{OTRD}), which considers future vessel trajectory trends to enhance similarity accuracy.

    \item We present an efficient \underline{\textbf{c}}ontinuous \underline{\textbf{s}}imilar \underline{\textbf{t}}rajectory \underline{\textbf{s}}earch (\texttt{CSTS}) algorithm, that integrates strategies to reduce unnecessary computations and optimize the search process.

    \item Experiments on two large real-world AIS datasets offer evidence that \texttt{ACTIVE} is capable of outperforming state-of-the-art methods by up to 70\% in terms of query time, 60\% in terms of hit rate, and reduces index construction costs significantly.
\end{itemize}

The rest of the paper is organized as follows. 
Section~\ref{sec:preliminary} presents preliminaries.
%
%
Section~\ref{sec:similarity} details the new similarity measure.
Section~\ref{sec:algorithm} presents the algorithms for continuous similar vessel trajectory search.
Section~\ref{sec:exp} reports the experimental results.
Section~\ref{sec:related} reviews related works. Section~\ref{sec:conclusion} concludes the paper and offers future directions.
\section{Preliminaries}
\label{sec:preliminary}

Section~\ref{ssec:trajectory} presents key definitions. Section~\ref{ssec:problem} defines the research problem. Section~\ref{ssec:framework} gives an overview of the \texttt{ACTIVE}.
Frequently used notation is listed in Table~\ref{tab:notation}.

\begin{table}[!htbp]
\centering
\footnotesize
{\setlength\tabcolsep{4pt}
\caption{Notation.}
\label{tab:notation}
\begin{tabular}{c l}
\Xhline{1pt}
\textbf{Symbol} & \textbf{Description} \\ \hline
$T_i$ & A trajectory \\
$s^u_i$ & The $u^{\textit{th}}$ segment of trajectory $T_i$ \\
$\mathcal{T}$ 
& A set of trajectories 
\\
$T_i = \langle s^1_i, ..., s^u_i\rangle$ & A list trajectory segments\\
$o$ & A moving object \\
$\mathcal{D}(o, T_i)$ & \makecell[l]{The distance between a moving object $o$\\ and a trajectory $T_i$} \\
$T_i^*$ & The trajectory $T_i$ with the last element removed \\
$d(p^{j'}, p^j)$ & The Euclidean distance between points $p^{j'}$ and $p^j$ \\
$L(p^{j'}, p^j)$ & The line segment between points $p^{j'}$ and $p^j$ \\
\Xhline{1pt}
\end{tabular}}
\end{table}

\vspace*{-7pt}
\subsection{Trajectory}
\label{ssec:trajectory}

\begin{definition}[GPS Point]
A GPS point $p=(x, y, t)$ includes a location with longitude $x$ and latitude $y$ and timestamp $t$.
\end{definition}

For simplicity, we refer to a GPS point as a point $(x, y)$ when it is in a snapshot at a timestamp.

\begin{figure*}[!htbp]
    \centering
    \includegraphics[width=\textwidth]{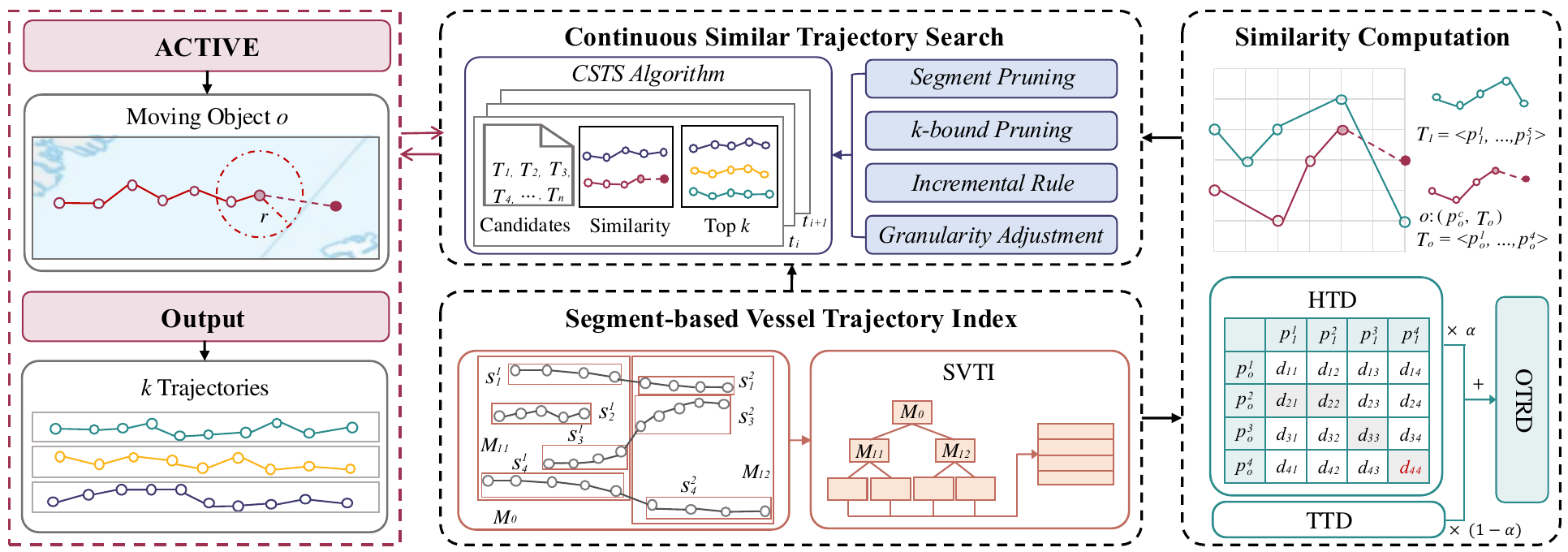}
    \caption{\texttt{ACTIVE} framework.}
    \label{fig:framework}
    \vspace{-5mm}
\end{figure*}

\begin{definition}[Historical Trajectory]
A \emph{historical trajectory} $T_i$ is a finite, time-ordered sequence $T_i = \langle p_i^1, p_i^2, ..., p_i^n \rangle$. Next, $\mathcal{T} = \{T_i \mid 1 \leq i \leq |\mathcal{T}|\}$ is a set of $|\mathcal{T}|$ historical trajectories.
\end{definition}

In Fig.~\ref{fig:trajectory}, $\mathcal{T} = \{T_1, T_2, T_3, T_4, T_5\}$. We simplify $T_i$ to $T$ when the index $i$ is not relevant. 


\begin{definition}[Segment]\label{def:segment}
A \emph{segment} $s_i = \langle p_i^{j'}, \dots, p_i^j \rangle$ of a trajectory $T_i$ is a continuous sub-sequence of $T_i$ defined by an index range $[j', j]$, where $1 \leq j' \leq j \leq n$. 
\end{definition}

Building on Definition~\ref{def:segment}, a trajectory $T_i$ can be represented as a sequence of segments $T_i = \langle s_i^1, s_i^2, \dots, s_i^u \rangle$, where $s_i^u$ is the $u^{\textit{th}}$ segment of $T_i$. For example, in Fig.~\ref{fig:trajectory}, trajectory $T_1$ consists of two segments, $s_1^1$ and $s_1^2$, thus $T_1 = \langle s_1^1, s_1^2 \rangle$, where $s_1^1=\langle p_1^1, p_1^2, p_1^3, p_1^4 \rangle$ and $s_1^2=\langle p_1^5, p_1^6, p_1^7, p_1^8 \rangle$.


\begin{definition}[Moving Object]
A \emph{moving object} $o$ changes its position over time in a given space. The destination of an object $o$ is given by $p_o^d$. Next, $o_t= (p_o^c, T_o)$ represents $o$'s information at  time $t$, where $p_o^c$ is the position of $o$ at time $t$ and $T_o = \langle p_o^1, p_o^2, \dots, p_o^c \rangle$ is the trajectory of $o$ up to time $t$.
\end{definition}



\begin{example}\label{ex:moving_object}
In Fig.~\ref{fig:trajectory}, at time $t$, $o_t = (p_o^6, \langle p_o^1, p_o^2, p_o^3, p_o^4,$ $p_o^5, p_o^6 \rangle)$, which means its current position is $p_o^6$ and its trajectory up to $t$ is $\langle p_o^1, p_o^2, p_o^3, p_o^4,$ $p_o^5, p_o^6 \rangle$. 
\end{example}





\begin{definition}[Pivotal Point]
Given a historical trajectory $T = \langle p^1, p^2, \dots, p^n \rangle$ and a moving object $o$ with $o_t = (p_o^c, T_o)$, the \emph{pivotal point} $p^x$ of $T$ w.r.t. $o$ is the point in $T$ closest to $p_o^c$ at time $t$. Letting $p^x=p^j$, the original trajectory $T$ is represented as $(T', T^L)$, where $T' = \langle p^1, p^2, \dots, p^j \rangle$ and $T^L = L(p^x, p^n)$ is the line segment between pivotal point $p^x$ and the last point $p^n$ in $T$.
\end{definition}

\begin{example}\label{ex:pivotal}
Continuing Example~\ref{ex:moving_object}, the pivotal point $p_4^x$ of $T_4$ w.r.t. $o$ is $p_4^5$, as it is the closest to $p_o^6$. Consequently, we use $(T_4', T_4^L)$ to represent $T_4$, where $T_4' = \langle p_4^1, p_4^2, p_4^3, p_4^4, p_4^5 \rangle$ and $T_4^L = L(p_4^5, p_4^7)$ is the dotted line between $p_4^5$ and $p_4^7$.
\end{example}

\begin{definition}[Range Query]
Given a range $r$, a moving object $o$ with $o_t = (p_o^c, T_o)$, a range query $\mathcal{R}(p_o^c, r)$ returns the set of historical trajectories $\mathcal{T}_r = \{T_i \mid T_i \in \mathcal{T} \wedge d(p_o^c, p_i^x) \leq r \}$, where $p_i^x$ is the pivotal point of trajectory $T_i$, and $d(p_o^c, p_i^x)$ is the distance between the current position $p_o^c$ of $o$ and the pivotal point $p_i^x$ of $T_i$. 
\end{definition}


\begin{example}\label{ex:range_query}
Continuing Example~\ref{ex:pivotal}, given the range $r$ in the example, the range query $\mathcal{R}(p_o^c, r)$ returns $\mathcal{T}_r = \{T_1, T_3, T_4\}$.
\end{example}

\vspace*{-7pt}
\subsection{Problem Definition}
\label{ssec:problem}
\begin{research}[Continuous Similar Trajectory Search (\texttt{CSTS})]
Given a moving object $o$, a set of historical trajectories $\mathcal{T}$, a range $r$, an integer $k$, and a similarity measure $\mathcal{D}(\cdot, \cdot)$, \texttt{CSTS}$(o, \mathcal{T}, r, k)$ continuously identifies the $k$ most similar trajectories $\Theta_t$ for $o$ at each timestamp $t$, where $o_t = (p_o^c, T_o)$, satisfying $\forall T \in \Theta_t\,  (T \in \mathcal{T}_r) \wedge \forall T' \notin \Theta_t\,  (\mathcal{D}(o, T') \geq \mathcal{D}(o, T))$.
\end{research}

Here, $\mathcal{D}(o, T)$ is the similarity measure \texttt{OTRD} (defined in Section~\ref{ssec:traj_dist}). Continuing Example~\ref{ex:range_query}, given $k=2$, $\Theta_t = \{T_1, T_4\}$.

\vspace*{-5pt}
\subsection{Framework Overview}
\label{ssec:framework}
Fig.~\ref{fig:framework} shows the three key modules in \texttt{ACTIVE} framework:


\noindent \textbf{Similarity Computation module.} It computes the real-time distance between a moving object and historical trajectories using the novel similarity measure \texttt{OTRD}, which consists of the Historical Trajectory Distance (HTD) and the Target-Trajectory Distance (TTD). The \texttt{OTRD} measure prioritizes the future trends of vessel movements and the target destination while also considering the relevance of trajectory segments and the weighted significance of individual points. 

\noindent \textbf{Segment-based Vessel Trajectory Index module.} Historical trajectories are partitioned into smaller segments using an MBR-optimization method and organized in an index. This index captures connection information, enabling efficient reconstruction of full trajectories when needed and supports spatially optimized search. 

\noindent \textbf{Continuous Similar Trajectory Search module.} Using the proposed \texttt{CSTS} algorithm, this module performs continuous search with four speed-up strategies that exploit geometric properties, utilize intermediate data, such as similarities computed at previous timestamps, and adjust trajectory granularity to reduce computational cost while preserving accuracy. Given a moving object $o$, a set of historical trajectories $\mathcal{T}$, a search range $r$, and an integer $k$, it continuously identifies the $k$ most similar trajectories $\Theta_t$ for $o$ at each timestamp $t$. 

\section{Similarity Measure}
\label{sec:similarity}
\subsection{Case Analysis}
\label{subsec:case}
Existing trajectory distance measures, such as Dynamic Time Warping (DTW), Hausdorff distance, and Fr{\'e}chet distance, typically evaluate the entire trajectory and treat each point with equal significance. However, these methods exhibit certain limitations when applied to real-time trajectory distance computations. Consider the following cases:

\begin{figure*}[ht]
    \centering
    \includegraphics[width=0.8\textwidth]{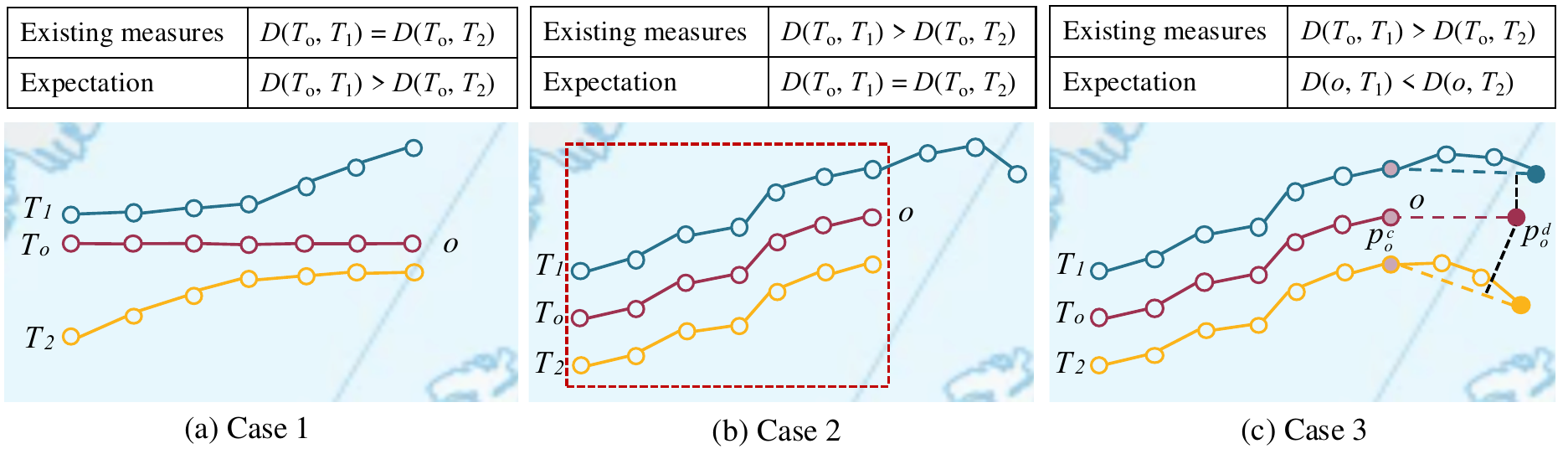}
    \caption{Examples of gaps between existing similarity measures and real needs.} \label{fig:cases}
    \vspace{-5mm}
\end{figure*}

\noindent \textbf{Case 1: Trend Neglect.} Traditional trajectory similarity measures  focus on minimizing geometric or temporal differences without considering the directional movement. In Fig.~\ref{fig:cases}(a), the distance between trajectories $T_1$ and $T_o$ is computed to be the same as that between $T_2$ and $T_o$
under existing measures. However, in many real-world applications, such as predictive modeling or motion forecasting, the trajectory trend plays a critical role in determining similarity. For example, although $T_2$ is farther from $T_o$ at earlier timestamps, its trend more closely aligns with $T_o$’s current movement direction, suggesting that $T_2$ is more likely to converge with $T_o$ in the near future. This indicates that ignoring the trend can lead to suboptimal or misleading results, particularly in scenarios where the future behavior of trajectories is of primary interest.

\noindent \textbf{Case 2: Segment Relevance.} Conventional trajectory similarity measures assume that all segments or points along a trajectory contribute equally to the similarity computation. This assumption overlooks the fact that, in many downstream tasks, certain segments of a trajectory may carry more relevance than others. For example, in Fig.~\ref{fig:cases}(b), given a trajectory $T_o$ and a set of trajectories ${T_1, T_2}$, only the segment highlighted in the red box is of primary interest for comparison. Ignoring the varying importance of trajectory segments can lead to suboptimal decision-making in critical scenarios.

\noindent \textbf{Case 3: Target Point Neglect.} In many practical applications, such as logistics and real-time navigation, the target destination of a moving object is of paramount importance. Current trajectory similarity measures, however, are predominantly backward-looking, focusing on historical data while disregarding the significance of the target and the trajectory’s future course. 
For instance, as shown in Fig.~\ref{fig:cases}(c), while the historical paths of trajectories $T_1$ and $T_2$ are similar, $T_1$' trend after the pivotal point is more aligned with the projected path of $T_o$ when the target destination is taken into consideration. 
Incorporating the target point into trajectory similarity measures would not only enhance the accuracy of real-time trajectory search but also enable more reliable predictions of future movements in applications such as autonomous navigation, shipping logistics, and traffic management.

\begin{figure}[!htbp]
    \centering
    \includegraphics[width=\columnwidth]{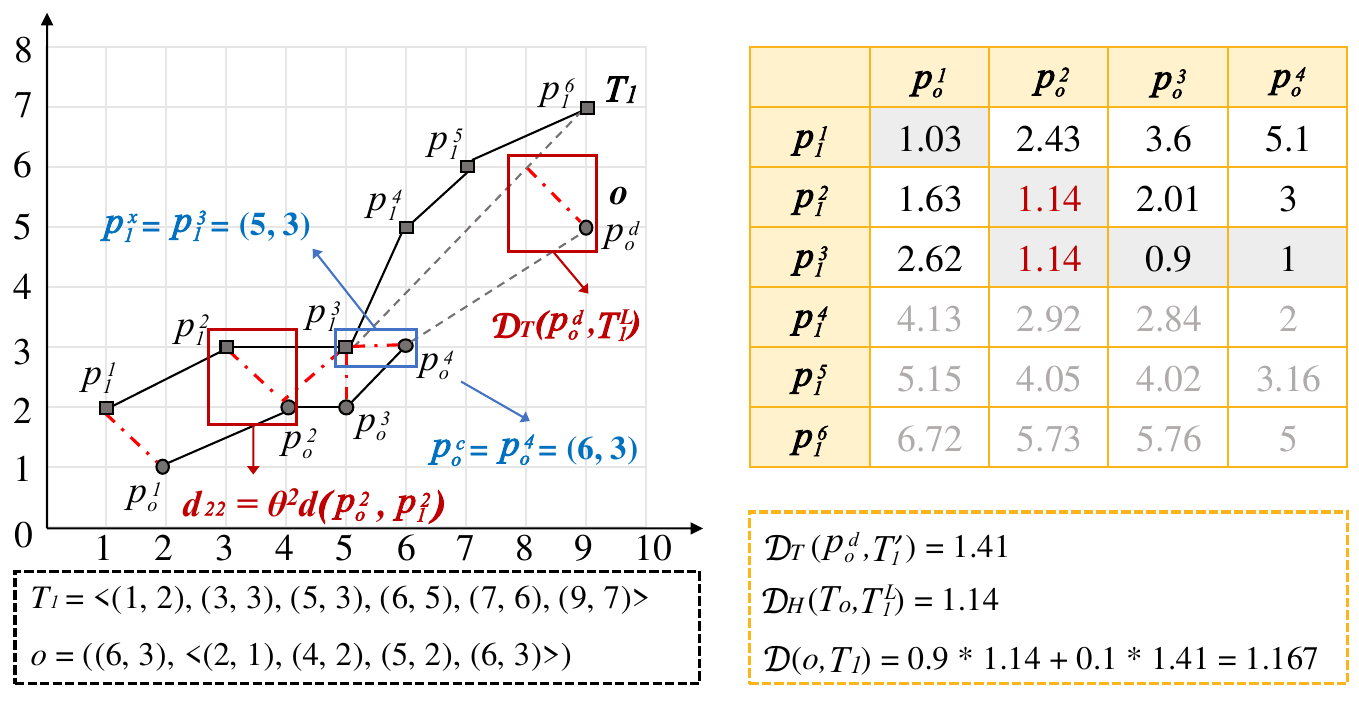}
    \caption{Example of \texttt{OTRD} measure.}
    \label{fig:OTRD}
    \vspace{-5mm}
\end{figure}

\subsection{Object-Trajectory Real-time Distance}
\label{ssec:traj_dist}

The \underline{\textbf{o}}bject-\underline{\textbf{t}}rajectory \underline{\textbf{r}}eal-time \underline{\textbf{d}}istance (\texttt{OTRD}) is a similarity measure function $\mathcal{D}(o, T_i)$ to measure the distance between a moving object $o$ at timestamp $t$ with information $(p_o^c, T_o)$ and a trajectory $T_i$. The distance between $o$ and $T_i$ consists of the Historical Trajectory Distance (HTD) and the Target-trajectory Distance (TTD).
Specifically, the HTD is the distance between $T_o$ of $o$ and $T'_i$ of $T_i$, denoted as $\mathcal{D}_H(T_o, T'_i)$, where $T'_i = \langle p_i^1, \dots, p_i^x \rangle$.
The TTD is the distance from $o$' target destination $p_o^d$ to the line $T^L = L(p_i^x, p_i^n)$, denoted as $\mathcal{D}_T(p_o^d, T_i^L)$. 
The \texttt{OTRD} considers more about the points near the current position of the moving object. Besides, the target destination is also considered to improve the accuracy of the similarity in terms of the future trend. 

\begin{definition}[Object-Trajectory Real-time Distance]
Given a \emph{moving object} $o$ at timestamp $t$ with $o_t = (p_o^c, T_o)$ and a trajectory $T_i = \langle p_i^1, p_i^2, \dots, p_i^n\rangle$, the Object-Trajectory Real-time Distance between $o$ and $T_i$ is defined as follows:
\begin{equation}\label{equ: OTRD}
\mathcal{D}(o, T_i) = \alpha \cdot \mathcal{D}_H(T_o, T'_i) + (1 - \alpha) \cdot \mathcal{D}_T(p_o^d, T_i^L).
\end{equation}

\begin{equation}\label{equ: HTD}
\begin{aligned}
\mathcal{D}_H(T_o, T'_i) = 
\begin{cases}
\theta^{|T''_o|-1} d(p_o^*, p_i^*),  \mbox{ if } |T_o| = 1 \mbox{ and } |T'_i| = 1, \\
\textit{min}\{\theta^{|T''_o|-|T_o|}d(p_o^*, p_i^*), \mathcal{D}_H(T_o^*, T'_i)\}, \\\mbox{ } \mbox{ } \mbox{ } \mbox{ } \mbox{ } \mbox{ } \mbox{ if } |T_o| > 1 \mbox{ and } |T'_i| = 1, \\
\textit{max}\{\mathcal{D}_H(T''_o, {T'_i}^*), \textit{min}\{\theta^{|T''_o|-|T_o|}d(p_o^*, p_i^*), \\\mbox{ } \mbox{ } \mbox{ } \mbox{ } \mbox{ } \mbox{ } \mathcal{D}_H(T_o^*, T'_i)\}\}, \\\mbox{ } \mbox{ } \mbox{ } \mbox{ } \mbox{ } \mbox{ } \mbox{ if } |T_o| > 1 \mbox{ and } |T'_i| > 1.
\end{cases}
\end{aligned}
\end{equation}

\begin{equation}\label{equ:TTD}
\begin{aligned}
\mathcal{D}_T(p_o^d, T_i^L) = d(p_o^d, L(p_i^x, p_i^n)).
\end{aligned}
\end{equation}
\end{definition}
Here, $d(p_o^*, p_i^*)$ is the euclidean distance between $p_o^*$ and $p_i^*$, where $p_o^*$ and $p_i^*$ are last points in $T_o$ and $T'_i$. $\theta \in [0, 1]$ is the decay factor.
$T''_o$ is the original $T_o$.
$|T|$ is the number of points in trajectory $T$.
$x = \text{argmin}_{0 < j \leq n} d (p_o^c, p_i^j)$ is the ID of the pivotal point in $T_i$. 
$T_o^*$ and ${T'_i}^*$ are the segments of $T_o$ and $T'_i$ with the last point removed.
The distance can be flexibly customized by a trade-off parameter $\alpha \in [0, 1]$ according to specific application needs. The process of \texttt{OTRD} computation is detailed in Section~\ref{ssec:similarity_computation}.

\begin{example}
Fig.~\ref{fig:OTRD} shows an example of \texttt{OTRD}. Given a trajectory $T_1 = \langle (1,2), (3,3), (5,3), (6,5), (7,6), (9,7) \rangle$, a moving object $o$ at time $t$ with the target destination $p_o^d = (9, 5)$ and $(p_o^c, T_o)$, where $p_o^c = (6, 3)$ and $T_o = \langle (2,1), (4,2), (5,2), (6,3) \rangle$, the trade-off parameter $\alpha = 0.9$, and the decay factor $\theta = 0.9$, we have the pivotal point of $T_1$, $p_1^x = p_1^3 = (5, 3)$, and two parts of $T_1$, $T'_1 = \langle (1,2), (3,3), (5,3) \rangle$ and $T_1^L = L((5,3), (9, 7))$. The HTD and TTD can be computed accordingly, i.e.,
$\mathcal{D}_H(T_o, T'_1) = 1.14$ and $\mathcal{D}_T(p_o^d, T_1^L) = 1.41$. Therefore, the \texttt{OTRD} between $o$ and $T_1$ is $\mathcal{D}(o, T_1) = 0.9 * 1.14 + 0.1 * 1.41 = 1.167$.
\end{example}

\section{Search Algorithm}
\label{sec:algorithm}

In this section, we present an efficient search algorithm for \texttt{CSTS}, which operates on a segment-based trajectory index and incorporates four acceleration strategies.

\subsection{Segment-based Trajectory Index}
\label{sec:index}


The \underline{\textbf{s}}egment-based \underline{\textbf{v}}essel \underline{\textbf{t}}rajectory \underline{\textbf{i}}ndex (\texttt{SVTI}) is a spatial access method designed to index vessel trajectory data efficiently. Each trajectory is divided into segments and an R-Tree structure organizes these segments hierarchically, enabling fast querying by grouping spatially close segments. 
Each leaf node stores the Minimum Bounding Rectangle (MBR) of its segments and a set of tuples, such as ($i, U$), where $i$ is the trajectory ID, and $U$ is the set of segment IDs within $T_i$. The leaf nodes also link to detailed information, including the sequence of points forming each segment and the search path for preceding segments, enabling seamless segment-to-trajectory mapping. Non-leaf nodes store the MBR of the node itself and references to trajectory and segment IDs in their child nodes. This hierarchical design preserves spatial relationships and trajectory continuity across different tree levels, ensuring efficient indexing and querying.
\begin{figure*}[!htbp]
    \centering
    \includegraphics[width=0.85\textwidth]{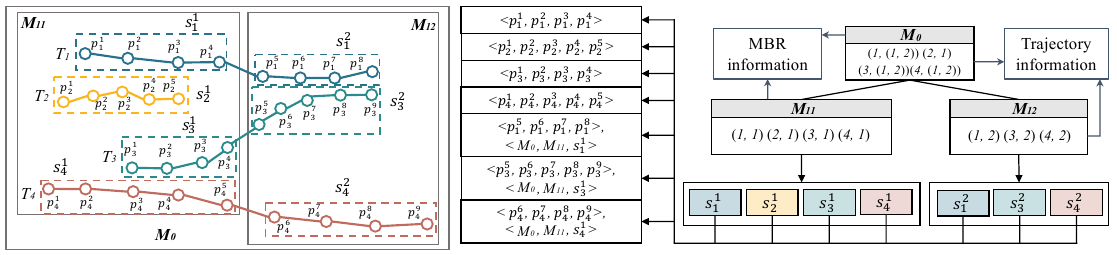}
    \caption{An example of \texttt{SVTI}.}
    \label{fig:index}
    \vspace{-5mm}
\end{figure*}
\begin{example}
Fig.~\ref{fig:index} illustrates an example of \texttt{SVTI}. The index includes two leaf nodes, $M_{11}$ and $M_{12}$. In $M_{11}$, four segments are stored, each corresponding to one trajectory: $T_1$, $T_2$, $T_3$, $T_4$. $M_{12}$ contains three segments, involving trajectories $T_1$, $T_3$, $T_4$. Each leaf node is linked to both the segment data and the connection information. For example, $M_{12}$ contains $s^2_1$, which is the second segment in $T_1$. Detailed segment information is associated with $s^2_1$, including its preceding segment $s^1_1$ and the path $\langle M_0, M_{11} \rangle$ that leads to $s^1_1$. The two leaf nodes, $M_{11}$ and $M_{12}$, are then grouped into a higher-level node $M_0$, which serves as the root node in this case. $M_0$ manages a total of seven segments, involving four different trajectories.
\end{example}


The performance of the index is significantly influenced by both the number of segments and the total area covered by the MBRs. Reducing the total MBR area minimizes overlap between nodes, thereby improving the efficiency of spatial indexing and query performance.
Therefore, the goal of segment partitioning is to split a trajectory such that the cumulative area of the MBRs for the resulting segments is minimized.

The segment partitioning method uses dynamic programming to determine the optimal segmentation. For each point $p^j$ in a given trajectory, the minimum total MBR area, $F[j]$, is computed by considering all possible previous points, starting from $p^{j - l + 1}$, where $l_\textit{min} \leq l \leq l_\textit{max}$. Here, $l_\textit{min}$ and $l_\textit{max}$ denote the minimum and maximum segment lengths, respectively. The recurrence relation is defined as follows:
\begin{equation}\label{equ: segment}
\begin{aligned}
F[j] = \textit{min}\{F[j - l] + \mathcal{M}_{\textit{area}}(p^{j - l + 1}, p^j)| l_\textit{min} \leq l \leq l_\textit{max}\},  
\end{aligned}
\end{equation}
where $\mathcal{M}_{\textit{area}}(p^{j - l + 1}, p^j)$ is the MBR area of segment $\langle p^{j - l + 1}, ..., p^j \rangle$.
Parameter $l_\textit{min}$ plays a critical role in controlling the number of segments, which directly impacts the construction of the R-tree-based index structure. By reducing the number of segments, the efficiency of tree indexing and query processing is improved.
Meanwhile, $l_\textit{max}$ defines an upper bound on the time complexity of pruning operations. This ensures that the computational cost remains manageable, even as the trajectory data becomes more complex.
By balancing segment length constraints and MBR optimization, the partitioning process enhances query performance and overall system efficiency in managing large-scale trajectory data.
\begin{algorithm}
    \caption{\textsc{Segment\_partitioning}($\mathcal{T}$, $l_\textit{max}$, $l_\textit{min}$)} \label{alg:segment}
    \begin{algorithmic}[1]
        \Statex \textbf{Input:} A vessel trajectory set $\mathcal{T}$, the maximum segment length $l_\textit{max}$, and the minimum segment length $l_\textit{min}$.
        \Statex \textbf{Output:} A segment set $\mathcal{S}$.
        \State Initialize a segment set $\mathcal{S}$ $\gets$ $\emptyset$
        \For{$T_i = \langle p_i^1, p_i^2, ..., p_i^n \rangle$ in $\mathcal{T}$}
            \State Initialize a list for $T_i$'s segment set $S_i$ $\gets$ $\emptyset$ 
            \State Initialize a list for the total MBR area $F$ $\gets$ $\emptyset$ 
            \For{$j$ in $[1, n]$}
                \If{$j \leq l_\textit{min}$}
                    $F[j]$ $\gets$ $0$
                \Else { }
                    $F[j]$ $\gets$ $\infty$
                \EndIf
                \For{$l$ in $[l_\textit{min}, l_\textit{max}]$}
                    \If{$j - l \geq 0$}
                        \State $m_l$ $\gets$ $\mathcal{M}_{area}$($p_i^{j - l + 1}$, $p_i^j$)
                        \If{$j - l$ is $0$} 
                            $m$ $\gets$ $m_l$
                        \Else { }
                            $m$ $\gets$ $F[j - l] + m_l$
                        \EndIf
                        \If{$F[j] > m$}
                            \State $F[j]$ $\gets$ $m$
                            \State $S_i[j]$ $\gets$ $S_i[j - l]$ $\cup$ $\langle p_i^{j - l + 1}, ..., p_i^j \rangle$
                        \EndIf
                    \EndIf
                \EndFor
            \EndFor
            \State $\mathcal{S} \gets \mathcal{S} \cup S_i[n]$
            
        \EndFor
        \State \textbf{return} $\mathcal{S}$
    \end{algorithmic}
\end{algorithm}

Algorithm~\ref{alg:segment} outlines the segment partitioning process in detail. Initially, a segment set $\mathcal{S}$ is initialized (line~1). For each trajectory $T_i$, the algorithm identifies the optimal segmentation (lines~2--16). Specifically, a list is created for $T_i$'s segment set, where the $j$th item represents the current optimal segment set up to point $p^j$ (line~3). Additionally, another list is initialized to store the total MBR area, where the $j$th item represents the minimum total MBR area up to point $p^j$ (line~4). For each point $p^j$ in $T_i$, the minimum total MBR area up to $p^j$ is computed (lines~5--15). If $j$ is less than $l_\textit{min}$, the area is set to $0$; otherwise, it is initialized to $\infty$ (lines~6--7). For each possible starting point $p^{j - l + 1}$, where $l_\textit{min} \leq l \leq l_\textit{max}$, the MBR area of the segment $\langle p^{j - l +1}, ..., p^j \rangle$ is computed (lines~8--10). If the total MBR area is smaller than the current value, both the area and the segment set are updated (lines~11--15). Once all points in $T_i$ have been processed, the final segment set for $T_i$ is added to the overall segment set $\mathcal{S}$ (line~16). Ultimately, the algorithm returns the overall segment set $\mathcal{S}$ (line~17). The complexity of each trajectory $T_i$ partitioning is $O(n(l_{max}-l_{min}+1))$, where $n$ is the number of points in $T_i$.

\subsection{Speed-up Strategies}
\label{ssec:speed-up strategies}

To enhance the efficiency of similarity computation and continuous similar trajectory search, we propose four speed-up strategies. These strategies leverage geometric properties, intermediate data, and trajectory granularity adjustments to reduce computational overhead while preserving accuracy.

The similarity computation operates segmented trajectories to improve the efficiency. The MBR of a segment $s$ denoted as $\mathcal{M}(s) = (v_l^u, v_l^b, v_r^u, v_r^b)$, where $v_l^u$ and $v_l^d$ are the upper-left and the bottom-left vertices of the MBR, while $v_r^u$ and $v_r^d$ are the upper-right and the bottom-right vertices. 
For a point $p$ outside of $\mathcal{M}(s)$, the distance between $p$ and $\mathcal{M}(s)$ is computed as follows:
\begin{equation}
\begin{aligned}
\mathcal{D}_s(p, \mathcal{M}(s)) = \text{min}\{d^*(p, v_l^u), d^*(p, v_l^b), d^*(p, v_r^u), d^*(p, v_r^b)\},
\end{aligned}
\end{equation}
where $d^*(p, v) = \text{min}\{|p.x - v.x|, |p.y - v.y|\}$.

\begin{strategy}[Segment Pruning]\label{Strategy: S1}
When computing the HTD between the moving object $o$ at timestamp $t$ with $o_t = (p_o^c, T_o)$ and a trajectory $T_i$, a segment $s_i$ can be pruned if 
\begin{equation}
\begin{aligned}
\theta^{c-j}\mathcal{D}_s(p_o^j, \mathcal{M}(s_i)) > \textit{dist}_{\textit{min}},
\end{aligned}
\end{equation}
where $c$ is the ID of $o$'s the current point at timestamp $t$, $p_o^j$ is a point in $o$'s historical trajectory $T_o$, $T'_i$ is the sub-trajectory of $T_i$ from the starting point to the pivotal point, $s_i$ is a segment in $T'_i$, and $\textit{dist}_{\textit{min}}$ is the current minimum distance between $p_o^j$ and $T'_i$. 
\end{strategy}

\begin{lemma}
Given a segment $s$ and a point $p$, the distance between $p$ and any point $p^j$ in $s$ is no less than $\mathcal{D}_s(p, \mathcal{M}(s))$.
\end{lemma}

\begin{proof}
If $p^j$ lies inside the rectangle, the shortest possible path from $p$ to $p^j$ must pass through a boundary point because the shortest distance from an external point to a shape is to its boundary. This boundary point must be closer or equal to the nearest vertex $v_{\textit{min}}$.
If $p^j$ lies on an edge, it lies between two vertices of the rectangle. Since the distance between $p$ and any point on the edge is bounded by the distance from $p$ to the closest vertex, the distance $d(p, p^j) \geq d(p,v_{\textit{min}}) = \mathcal{D}_s(p, \mathcal{M}(s))$.
\end{proof}
\vspace*{-3pt}
\begin{strategy}[$k$-bound Pruning]\label{Strategy: S2}
Given the current $k$-th smallest Object-Trajectory Real-time Distance $\textit{dist}_k$ between a moving object $o$ at timestamp $t$ with $o_t = (p_o^c, T_o)$ and the computed historical trajectories, a trajectory $T_i$ can be pruned if
\begin{equation}
\begin{aligned}
\mathcal{D}^l(p_o^j, T'_i) \geq \frac{\textit{dist}_k - (1-\alpha)\mathcal{D}_T(p_o^d, T_i^L)}{\alpha \cdot \theta^{c-j}} ,
\end{aligned}
\end{equation}
where 
\begin{equation}
\begin{aligned}
\mathcal{D}^l(p_o^j, T'_i) =  \text{min}\{d(p_o^j, p_i)| p_i \in T'_i\}.
\end{aligned}
\end{equation}
\end{strategy}

\begin{lemma}\label{lemma: lower_bound}
Given a moving object $o$ at timestamp $t$ with $o_t = (p_o^c, T_o)$ and a trajectory $T_i$, the Historical Trajectory Distance $\mathcal{D}_H(T_o, T'_i)$ is no less than $d(p_o^c, p_i^x)$.
\end{lemma}

\begin{proof}
When computing HTD, it starts with the distance computation between the last points of $T_o$ and $T'_i$, i.e., $p_o^c$ and $p_i^x$. Therefore, $d(p_o^c, p_i^x)$ must in the results of $\textit{max}$ function in Equation~\ref{equ: HTD}. If there is an other value in $\textit{max}$ function results greater than $d(p_o^c, p_i^x)$, the final distance must greater than $d(p_o^c, p_i^x)$. Otherwise, $d(p_o^c, p_i^x)$ will be the final distance of $\mathcal{D}_H(T_o, T'_i)$.
\end{proof}

According to Lemma~\ref{lemma: lower_bound}, an early pruning strategy can be used before fully computing the distance between $o$ and $T_i$. $T_i$ can be pruned if $d(p_o^c, p_i^x) \geq (\textit{dist}_k - (1-\alpha) \cdot \mathcal{D}_T(p_o^d, T_i^L))/\alpha$.

To avoid re-computation, the intermediate information buffer during the search is maintained. It contains a list of information for each computed trajectory $T_i$, denoted as $B_I = (i, x', z', \textit{dist}'_H , y')$, where $i$ is the ID of $T_i$, $x'$ is the pivotal point ID in $T_i$ at previous timestamp $t'$, $z'$ is the segment ID of $p_i^{x'}$ in $T_i$, $\textit{dist}'_H$ is the HTD between $o$ and $T_i$ at previous timestamp $t'$, and $y'$ is the ID of $p_o^{y'}$ in $T_o$, where $\theta^{c - y'} * \text{min}\{d(p_o^{y'}, p_i)| p_i \in T'_i\} = \mathcal{D}_H(T_o, T'_i)$. 
Then, the search at timestamp $t$ can be accelerated by using the following incremental rule.

\begin{strategy}[Incremental Rule]\label{Strategy: S3}
Given the intermediate buffer of \texttt{CSTS}($o, \mathcal{T}, r, k$), if a trajectory $T_i$ in $\mathcal{T}$ has been computed in a previous timestamp, the similarity computation at timestamp $t$ can be accelerated by using the incremental rule in different cases instead of fully computing $\mathcal{D}(o, T_i)$.

\noindent\textbf{Case 1:} If the pivotal point ID in $T_i$ at $t$ is unchanged, then the TTD $\textit{dist}_T$ is unchanged and the HTD $\textit{dist}_H$ can be updated as follows:
\begin{equation}\label{equ:update}
\begin{aligned}
    \textit{dist}_H = \text{max}\{\theta \cdot \textit{dist}'_H, \text{min}\{d(p_o^c, p_i)| p_i \in T'_i\}\}.
\end{aligned}
\end{equation}

\noindent\textbf{Case 2:} If the pivotal point ID in $T_i$ at $t$ is greater than $x'$, then $\textit{dist}_T$ can be computed according to Equation~\ref{equ:TTD}.
%
%
If $\theta^{c-y'} \cdot \textit{min}\{d(p_o^{y'}, p_i)| p_i \in \langle p_i^{x'+1}, ..., p_i^x \rangle \} \geq \theta \cdot \textit{dist}'_H$, where $\langle p_i^{x'+1}, ..., p_i^x \rangle$ is the incremental part of $T'_i$, then $\textit{dist}_H$ can be updated as Equation~\ref{equ:update}.
%
Otherwise, compute $\textit{dist}_H$ according to Equation~\ref{equ: HTD}.

\noindent\textbf{Case 3:} If the pivotal point ID in $T_i$ at $t$ is less than $x'$,  $\textit{dist}_H$, and $\textit{dist}_T$ can be computed according to Equations~\ref{equ: HTD} and~\ref{equ:TTD}.

Finally, the \texttt{OTRD} can be updated according to Equation~\ref{equ: OTRD}.
\end{strategy}

To further speed up the search, we can adjust the granularity of each historical trajectory.
\begin{strategy}[Granularity Adjustment]\label{Strategy: S4}
Given a trajectory $T_i$ in $\mathcal{T}$, the pivotal point $p_i^x$ in $T_i$, and a granularity parameter $g$, we have a coarse-grained trajectory $T'_i(g) = \langle p_i^j, p_i^{j+g}, \dots, p_i^{x-g}, p_i^x \rangle$, where $1 \leq j \leq g$. Then, $T'_i(g)$ will be used in the HTD computation, i.e., $\mathcal{D}_H(T_o, T'_i(g))$.
\end{strategy}

However, this strategy may introduce errors to $\mathcal{D}_H(T_o, T'_i)$ since some points in $T_i$ are ignored. We deduce the lower and upper error bounds as follows:
\begin{equation}
\begin{aligned}
    \mathcal{E}^l = \mathcal{D}_H(T_o, T'_i) - d(p_o^c, p_i^x), 
\end{aligned}
\end{equation}
\begin{equation}
\begin{aligned}
    \mathcal{E}^u = \text{max}\{\theta^{c-j} \cdot d(p_o, p_i^j)|p_o \in T_o, p_i^j \in T'_i(g)\} - \mathcal{D}_H(T_o, T'_i). 
\end{aligned}
\end{equation}

The lower error bound $\mathcal{E}^l$ can be easily proved according to Lemma~\ref{lemma: lower_bound}. The upper error bound can be proved as follows.
\begin{proof}
If $\exists p_i^j \in T'_i \setminus T'_i(g)$, $\exists p_o \in T_o$, such that $\theta^{c-j} \cdot d(p_o, p_i^j) > \textit{max}\{\theta^{c-j} \cdot d(p_o, p_i^j)|p_o \in T_o, p_i^j \in T'_i(g)\}$, then $d(p_o, p_i^j)$ must be excluded in $\textit{min}$ function in Equation~\ref{equ: HTD}. 
Therefore, the upper bound of $\mathcal{D}_H(T_o, T'_i)$ is $\textit{max}\{\theta^{c-j} \cdot d(p_o, p_i^j)|p_o \in T_o, p_i^j \in T'_i(g)\}$.
\end{proof}

To apply the incremental rule to $T'_i(g)$, we design the update rule of $T'_i(g)$ at timestamp $t$ with the intermediate buffer $B_I$ when the pivotal point is changed from $p_i^{x'}$ to $p_i^x$.
\begin{equation}
\begin{aligned}
    T'_i(g) = 
    \begin{cases}
    \langle p_i^j, ..., p_i^{x-g}, p_i^{x} \rangle &\mbox{if } \gamma = 0,\\
    \langle p_i^j, ..., p_i^{x-\gamma-g}, p_i^{x-\gamma}, p_i^{x} \rangle &\mbox{if }\gamma > 0\mbox{ and } x > x',\\
    \langle p_i^j, ..., p_i^{x+\gamma - g}, p_i^{x} \rangle &\mbox{if }\gamma > 0\mbox{ and } x < x'.
    \end{cases}
\end{aligned}
\end{equation}
\begin{equation}
\begin{aligned}
    \gamma = |x-x'| \mbox{ mod } g.
\end{aligned}
\end{equation}

The proposed speed-up strategies can enhance the efficiency of continuous similar trajectory search by leveraging mathematical properties of trajectory segments and intermediate results. Segment pruning and 
$k$-bound pruning effectively eliminate irrelevant candidates early in the process, while the incremental rule optimizes computations through reuse of prior results. Granularity adjustment further reduces computational overhead by coarsening trajectory representation within controlled error bounds. Together, these strategies provide a robust framework to accelerate similarity computations, enabling real-time and scalable trajectory analysis.

\subsection{\texttt{OTRD} Computation}
\label{ssec:similarity_computation}
The similarity computation between a moving object $o$ and a trajectory $T_i$ is performed in several key steps. First, the algorithm identifies the pivotal point $p_i^x$ of $T_i$ to the current position $p_o^c$ of $o$. Next, it computes the distance between $o$' target destination $p_o^d$ and the line $L(p_i^x, p_i^n)$, where $p_i^n$ is the end point of $T_i$.
Afterward, the distance between the $o$’s current trajectory $T_o$ and the sub-trajectory $T'_i = \langle p_i^1,..., p_i^x \rangle$ from $T_i$. To enhance the efficiency during this process, the speed-up strategies are employed to avoid unnecessary distance computations. Finally, the overall similarity is determined by summing these distances with a trade-off parameter $\alpha$, providing a comprehensive measure of how closely the moving object’s trajectory aligns with the historical one. 

\begin{algorithm}
    \caption{\textsc{OTRD\_Computation}($o$, $T_i$, $\alpha$, $\theta$, $g$, $\textit{dist}_k$, $B_I$)} 
    \label{alg:similarity_computation}
    \small
    \begin{algorithmic}[1]
        \Statex \textbf{Input:} A moving object $o$, a trajectory $T_i$, the trade-off parameter $\alpha$, the decay parameter $\theta$, the granularity parameter $g$, the current top-$k$ threshold $\textit{dist}_k$, and the intermediate buffer $B_I$. 
        \Statex \textbf{Output:} The distance between $o$ and $T_i$, and the updated intermediate buffer $B_I$.
        \State $s_i^z \gets \textsc{Find\_Pivotal\_Segment}(p_o^c, T_i)$
        \State $p_i^x \gets \textsc{Find\_Pivotal\_Point}(p_o^c, T_i)$
        \State $\textit{isValid} \gets \textit{false}$, $p_\textit{max} \gets B_I[T_i].y'$
        \If{$B_I[T_i].x'$ < $x$}
            \State $\textit{dist}_n \gets$ \textsc{Min\_Dist}($B_I[T_i].y'$, $B_I[T_i].z'$, $z$, $T$, $B_I[T_i].\textit{dist}'_H$, $g$)
            \If{$\textit{dist}_n > B_I[T_i].\textit{dist}'_H$}
                $\textit{isValid} \gets \textit{true}$ 
            \EndIf
        \EndIf
        \State $\textit{dist}_T \gets \textsc{Compute\_Dist\_T}(p_o^d, p_i^x, s_i^z, T_i)$, $\textit{dist}_H \gets 0$
        \For {each point $p_o^j$ in $T_o$}
            \If {$p_o^j \neq p_o^c$ and $isValid$}
                continue \Comment{Strategy~\ref{Strategy: S3}}
            \EndIf
            \State $\textit{dist}_n \gets$ \textsc{Min\_Dist}($j$, $0$, $z$, $T$, $\infty$, $g$)
            \If {$\textit{dist}_H < \theta^{c-j} \cdot \textit{dist}_n$}
                \State $\textit{dist}_H \gets \theta^{c-j} \cdot \textit{dist}_n$, $p_{\textit{max}} \gets j$
                \State $\textit{dist}'_k \gets (\textit{dist}_k - (1 - \alpha)\cdot \textit{dist}_T)/(\alpha \cdot \theta^{c-j})$
                \If {$\textit{dist}_H > \textit{dist}'_k$}
                    \textbf{break} \Comment{Strategy~\ref{Strategy: S2}}
                \EndIf
            \EndIf
        \EndFor
        \State $\textit{dist} \gets \alpha \cdot \textit{dist}_H + (1 - \alpha) \cdot \textit{dist}_T$
        \State $B_I \gets$ \textsc{Update\_B}($T, x, z, \textit{dist}_H, p_{\textit{max}}$)
        \State \textbf{return} $\textit{dist}$ and $B_I$
        \Function{\textsc{Min\_Dist}}{$j$, $z'$, $z$, $T_i$, $\textit{dist}_n$, $g$}
            \For {each segment $s_i^u$ in $T_i$ with $u \in (z', z]$}
                \State $\textit{dist}_s \gets \textsc{Compute\_Dist\_S}(p_o^j, s_i^u)$  
                \If {$\textit{dist}_s > \textit{dist}_n$}
                    \textbf{continue} \Comment{Strategy~\ref{Strategy: S1}}
                \EndIf
                \For {each point $p_i$ in $s_i^u$ with $g$} \Comment{Strategy~\ref{Strategy: S4}}
                    \State $\textit{dist}_p \gets d(p_o^j, p_i)$, $\textit{dist}_n \gets \textit{min}(\textit{dist}_n, \textit{dist}_p)$
                \EndFor
            \EndFor
            \State \textbf{return} $\textit{dist}_n$
        \EndFunction
    \end{algorithmic}
\end{algorithm}
Algorithm~\ref{alg:similarity_computation} shows the details of the similarity computation.
Specifically, we first identify the nearest segment $s_i^z$ and the nearest point $p_i^x$ on the historical trajectory $T_i$ to the moving object's current position $p_o^c$ (lines~1--2). 
We initialize a boolean flag $\textit{isValid}$ to check if the pre-computed distance in the intermediate buffer $B_I$ is still valid (line~3). 
A variable $p_{\textit{max}}$ is also initialized to represent the ID of the point $p_o^y$ in $T_o$, where $\theta^{c - y} * \textit{min}\{d(p_o^y, p_i)| p_i \in T'_i\} = \mathcal{D}_H(T_o, T'_i)$ (line~3). 
If the pivotal point ID of $T_i$ increases and the distance between $p_o^{y'}$ and the incremental part is greater than the maintained distance, $\textit{isValid}$ is set to $\textit{true}$ (lines~4--6). 
Then, the TTD $\textit{dist}_T$ is computed, which is the distance between the moving object's target destination $p_o^d$ and a line segment on the historical trajectory. (line~7)
Additionally, the HTD $\textit{dist}_H$ is initialized as $0$ (line~7). 
Then, $\textit{dist}_H$ is computed (lines~8--14). 
If the maintained distance is valid, we only update the distance between $p_o^c$ and $T'_i$ (Case~2 in Strategy~\ref{Strategy: S3}). 
Otherwise, $\textit{dist}_H$ should be computed by traversing every point $p_o^j$ in $T_o$ (from $p_o^c$ to previous points) to compute the minimum distance between $p_o^j$ and $T'_i$ using \textsc{Min\_Dist} function, where the segment pruning (Strategy~\ref{Strategy: S1}) and the granularity adjustment (Strategy~\ref{Strategy: S4}) are used during the computation(lines~18--24). During the traverse, the $p_{\textit{max}}$ is updated (line~12) and the $k$-bound pruning (Strategy~\ref{Strategy: S2}) is used to eliminate irrelevant candidates early (lines~13--14). 
Then, the \texttt{OTRD} value is computed by summing $\textit{dist}_H$ and $\textit{dist}_T$ with a trade-off parameter $\alpha$ (line~15). 
Next, the intermediate buffer is updated (line~16). 
Finally, the \texttt{OTRD} value and the updated intermediate buffer are returned (line~17).

\subsection{\texttt{CSTS} Algorithm}
\label{ssec:CSTS_algorithm}
The \texttt{CSTS} algorithm leverages the index \texttt{SVTI}, the similarity measure \texttt{OTRD}, and four speed-up strategies to efficiently identify the top $k$ similar trajectories for a moving object $o$ at each timestamp $t$.

Algorithm~\ref{alg:CSTS} outlines the \texttt{CSTS} algorithm. Initially, two parameters are set: the decay parameter $\theta$, which controls the influence of past trajectory points on similarity computations, and the trade-off parameter $\alpha$, which balances the weights of the two components in \texttt{OTRD} (lines~1--2). Next, for each timestamp $t$, we first initialize an intermediate buffer $B_I$ to maintain intermediate results (Strategy~\ref{Strategy: S3}) and a min-heap result buffer $B_k$ to keep the computed similarity values for each historical trajectory at timestamp $t$ (lines~4--5). Afterwards, a range query is executed to find all candidates $\mathcal{T}_r$ using the index \texttt{SVTI} (line~6). For each candidate trajectory $T_i$, it computes the similarity between $T_i$ and $o$ (lines~8--9). The current $k$-th smallest similarity value $\textit{dist}_k$ is retrieved from $B_k$. The similarity is computed using the \textsc{OTRD\_Computation} function incorporating pruning strategies such as segment pruning and $k$-bound pruning. The resulting similarity value and trajectory ID are inserted into $B_k$ (line~10). After evaluating all candidate, the top $k$ trajectories with the smallest \texttt{OTRD} values are extracted from $B_k$ (line~11). Finally, the top $k$ results for the current timestamp $t$ are returned (line~12).

\begin{algorithm}
    \caption{\textsc{CSTS\_Algorithm}($o$, $\mathcal{T}$, $r$, $k$, $g$, $SVTI$)} 
    \label{alg:CSTS}
    \small
    \begin{algorithmic}[1]
        \Statex \textbf{Input:} A moving object $o$, a historical trajectory set $\mathcal{T}$, a range $r$, an integer $k$, the granularity parameter $g$, segment-based vessel trajectory index $SVTI$.
        \Statex \textbf{Output:} Top $k$ similar trajectories in $\mathcal{T}$ at each timestamp.

        \State Initialize a decay parameter $\theta$
        \State Initialize a trade-off parameter $\alpha$
        \For{each timestamp $t$}
            \State Initialize an intermediate buffer: $B_I \gets \emptyset$
            \State Initialize a min-heap result buffer: $B_k \gets \emptyset$
            \State $\mathcal{T}_r \gets \textsc{Range}(o, \mathcal{T}, r, k, SVTI)$
            \For{each candidate trajectory $T_i$ in $\mathcal{T}_r$}
                \State $\textit{dist}_k \gets B_k.\textit{pop}()$
                \State $\textit{dist} \gets$ \textsc{OTRD\_Computation}($o$, $T_i$, $\alpha$, $\theta$, $g$, $\textit{dist}_k$, $B_I$)
                \State Insert ($\textit{dist}$, $T_i$) into $B_k$
            \EndFor
            
            \State $\Theta \gets \textsc{Get\_K}(B_k)$
            \State \textbf{return} $\Theta$ at timestamp $t$
        \EndFor
    \end{algorithmic}
\end{algorithm}

\subsection{Complexity Analysis}
\label{ssec:CSTS_complexity}

\noindent \textbf{Complexity of \texttt{OTRD} Computation.}  
Let $n_o$ and $n_i$ denote the numbers of points in the moving object trajectory $T_o$ and a historical trajectory $T_i$.
Without optimization, the distance computation requires comparing all point pairs, giving a complexity of $\mathcal{O}(n_o \cdot n_i)$.
The \texttt{OTRD\_Computation} function employs four optimizations:
(1) Segment pruning reduces examined segments, effectively decreasing $n_i$.
(2) $k$-bound pruning terminates early when the distance exceeds the $k$-th smallest value, reducing $n_i$ or $n_o$.
(3) Buffer reuse avoids redundant computations for unchanged parts, further limiting $n_i$ or $n_o$.
(4) Granularity adjustment samples points with step size $g$, reducing point-level comparisons to roughly $n_i/g$. With these, the average complexity becomes:
$\mathcal{O}(n'_o \cdot n'_i/g)$, 
where $n'_o$ and $n'_i$ are the effective numbers of points after pruning and buffer reuse.  
In practice, $n'_o \ll n_o$ and $n'_i \ll n_i$, resulting in a substantial reduction in computation time.  

\noindent \textbf{Complexity of \texttt{CSTS} Algorithm.}  
Let $|\mathcal{T}_r|$ be the number of candidate trajectories retrieved from the \texttt{SVTI} index at each timestamp.
The range query cost is $\mathcal{O}(\log |\mathcal{T}| + |\mathcal{T}_r|)$, and each similarity computation costs $\mathcal{O}(n'_o \cdot n'_i / g)$.
Maintaining the top-$k$ heap adds $\mathcal{O}(|\mathcal{T}_r| \log k)$. Thus, per-timestamp complexity is:
$\mathcal{O}(\log |\mathcal{T}| + |\mathcal{T}_r| \cdot (n'_o \cdot n'_i / g + \log k))$.
In practice, pruning and indexing greatly reduce $n'_i$, $n'_o$, $|\mathcal{T}_r|$, enabling efficient real-time search.

\section{Experiments}
\label{sec:exp}

\subsection{Overall Experimental Settings}
\label{ssec:setting}
All experiments were conducted on a Linux server running Ubuntu with an Intel Xeon Icelake processor featuring 60 cores at 5.7 GHz and 240 GB of memory. The system is equipped with dual NVIDIA A10 GPUs, each providing 23 GB VRAM, for a total of 46 GB of GPU memory. All neural network models are implemented using PyTorch and trained on the GPU, and all index construction and search algorithms are implemented in Python (Code for \texttt{CSTS}\footnote{\url{https://github.com/hyLiu1994/ACTIVE} (Last access: 2025.10)}).

\begin{table}[htbp]
    \centering
    \caption{Dataset statistics.}
    \label{tab:dataset}
    \small
    \begin{tabular}{l c c c c}
    \Xhline{1pt}
    \textbf{Dataset} & \textbf{Size (GB)} & \textbf{Trajectories} & \textbf{Points} \\
    \hline
    \texttt{DK-AIS}   & 37.43 & 746,302 & 216,189,595  \\
    \texttt{US-AIS}   & 15.61 & 737,133 & 154,279,662  \\
    \Xhline{1pt}
    \end{tabular}
\end{table}

\noindent\textbf{Datasets.} We evaluate the proposed methods on the following datasets. Dataset statistics are reported in Table~\ref{tab:dataset}.
\setlength{\parsep}{0pt}
\begin{itemize}[itemsep=1pt, leftmargin=10pt]
\item \texttt{DK-AIS}\footnote{\url{http://aisdata.ais.dk/?prefix=2024/} (Last access: 2025.10)}: The dataset is collected from a shore-based AIS system operated by the Danish Maritime Authority. 

\item \texttt{US-AIS}\footnote{\url{https://coast.noaa.gov/htdata/CMSP/AISDataHandler/2023/index.html} (Last access: 2025.1)}:
The dataset is collected by the US Coast Guard through an onboard navigation safety device.
\end{itemize}
  
\noindent\textbf{Tasks.} We conduct extensive experiments to address and analyze the following research questions (RQs):
\setlength{\parsep}{0pt}
\begin{itemize}[itemsep=1pt, leftmargin=10pt]

\item \textit{RQ1. Evaluation on \texttt{CSTS}:} How does the search algorithm \texttt{CSTS} perform in terms of efficiency and effectiveness?
    
\item \textit{RQ2. Ablation Study:} What is the impact of the speed-up strategies on overall performance?
\end{itemize}
The evaluation on \texttt{SVTI} could be found in Appendix.\\
\noindent\textbf{Metrics.} The metrics for each task are listed in Table~\ref{tab:baselines}.
To evaluate the performance of \texttt{CSTS}, we select ten query instances, execute the selected queries ten times, and report the average. For effectiveness evaluation, we employ the hit rate  ($\mathit{Hit}$):

\begin{equation}
  \begin{aligned}
  \mathit{Hit} = \frac{\lvert \mathcal{T}_f \cap \mathcal{T} \rvert}{k},
  \end{aligned}
\end{equation}

\noindent where $\mathcal{T}_f$ is the set of trajectories that cover the $k$-nearest neighbor ($k$NN) points of moving object $o$'s next future point and $\mathcal{T}$ is the query result.

\begin{table*}[!htbp]
    \caption{Metrics and baselines.}\label{tab:baselines}
    \small
    \centering
    \begin{tabular}{c c c c}
    \Xhline{1pt}
    \textbf{Questions} & \textbf{Metrics} & \textbf{Baselines}                                    
    \\ \hline 
      RQ1  & Query Time (ms), Hit (\%) & \texttt{O-PQT}, \texttt{O-PRT}, \texttt{H/F-TrajI}, \texttt{H/F-DFT}, \texttt{TrajCL-TI}, \texttt{CLEAR-TI} \\ 
      RQ2  & Query Time (ms), Hit (\%) & \makecell[c]{ \texttt{CSTS}$\setminus$S1, \texttt{CSTS}$\setminus$S2, \texttt{CSTS}$\setminus$S3, \texttt{CSTS}$\setminus$(S1+S2),\\ \texttt{CSTS}$\setminus$(S2+S3), \texttt{CSTS}$\setminus$(S1+S3), \texttt{CSTS}$\setminus$(S1+S2+S3)}   \\ \Xhline{1pt}
    \end{tabular}
    \vspace{-5mm}
\end{table*}

\noindent\textbf{Baselines.}
We compare the proposed methods with several baseline methods, as outlined in Table~\ref{tab:baselines}. 
Specifically, 
\texttt{CSTS} is compared with six traditional methods and two learning-based methods to evaluate its efficiency and effectiveness.
The baseline methods are listed as follows:
\begin{itemize}[itemsep=1pt, leftmargin=10pt]
\item \texttt{O-PQT}: \texttt{OTRD} + \texttt{PQT}. A continuous similar trajectory search method on a Point-based Quad-Tree (\texttt{PQT}) using the \texttt{OTRD} measure.

\item \texttt{O-PRT}: \texttt{OTRD} + \texttt{PRT}. A continuous similar trajectory search method on a Point-based R-Tree (\texttt{PRT}) using the \texttt{OTRD} measure.

\item \texttt{H-TrajI}/\texttt{F-TrajI}: Hausdorff/Fr{\'e}chet + Traj\_Index~\cite{xie2017distributed}. An R-Tree-based similar trajectory method using Hausdorff or Fr{\'e}chet distance, where the index is built according to the centroid of each data trajectory’s MBR.

\item \texttt{H-DFT}/\texttt{F-DFT}: Hausdorff/Fr{\'e}chet +  \texttt{DFT}~\cite{xie2017distributed}. A Distributed Framework for Trajectory similarity search (\texttt{DFT}) using Hausdorff or Fr{\'e}chet distance.

\item \texttt{TrajCL-TI}: \texttt{TrajCL}~\cite{chang2023contrastive} + \texttt{SVTI}. A trajectory similarity search method based on \texttt{SVTI}, employing a graph-based long-term dependency model for similarity computation.

\item \texttt{CLEAR-TI}: \texttt{CLEAR}~\cite{li2024clear} + \texttt{SVTI}. A trajectory similarity search method using \texttt{SVTI} with a similarity computation technique based on contrastive representation learning.
\end{itemize}

Additionally, we analyze the impact of different strategies on \texttt{CSTS}, as detailed in Table~\ref{tab:baselines}. For example, \texttt{CSTS}$\setminus$S1 represents \texttt{CSTS} without Strategy~\ref{Strategy: S1}, while \texttt{CSTS}$\setminus$(S1+S2+S3) indicates the removal of all strategies. By default, Strategy~\ref{Strategy: S4} is not applied. To explore the impact of Strategy~\ref{Strategy: S4}, we evaluate its effectiveness by varying the granularity parameter $g$.
\begin{table}[!htbp]
\caption{Parameter settings.}\label{tab:parameter_setting}
\small
\begin{tabular}{c c c}
\Xhline{1pt}
\textbf{Parameters} & \textbf{Questions} & \textbf{Settings}           \\ \hline
$\Lambda$ & RQ1 & D-1, D-2, D-3, D-4, \textbf{D-5} \\ 
$l_q$& RQ1 & \textbf{10}, 20, 30, 40, 50 \\ 
$r$  & RQ1 &   5, 10, \textbf{15}, 20, 25                           \\ 
$k$   & RQ1 & 30, 40, \textbf{50}, 60, 70  \\ 
$\theta$ & RQ1 & 0.45, \textbf{0.5}, 0.55, 0.6, 0.65 \\ 
$\alpha$ & RQ1 & 0.45, \textbf{0.5}, 0.55, 0.6, 0.65 \\ 
$g$ & RQ2 & \textbf{1}, 5, 10, 15, 20 \\
\Xhline{1pt}
\end{tabular}
\end{table}

\noindent\textbf{Parameters.}
The parameter settings are summarized in Table~\ref{tab:parameter_setting}, with default values in bold. A description of each parameter is provided below:
\begin{itemize}[itemsep=1pt, leftmargin=10pt]
    \item Data size ($\Lambda$): The number of the points in the AIS data. Five different data sizes are evaluated: D-1, D-2, D-3, D-4, and D-5, corresponding to approximately $20\%$, $40\%$, $60\%$, $80\%$, and $100\%$ of the original dataset.
    \item Query length ($l_q$): The length of the historical trajectory of moving object $o$ at the query's starting time $t$.
    \item Range ($r$): The range of the moving object $o$'s current position at each timestamp $t$. This is represented by the candidate rate, e.g., a range of 10 indicates that the candidate number is approximately $10 \cdot k$.
    \item Top $k$: The number of similar trajectories returned. 
    \item Decay factor ($\theta$): The decay factor used in the object-trajectory real-time distance computation.
    \item Trade-off factor ($\alpha$): The trade-off factor in the object-trajectory real-time distance computation. 
    \item Granularity ($g$): The granularity used in Strategy~\ref{Strategy: S4}.
\end{itemize}
For each query, we perform a continuous similar trajectory search across 20 timestamps and report the average results.

\subsection{Evaluation on \texttt{CSTS}}
\label{ssec:efficiency_csts}

We investigate the query time and hit rate of \texttt{CSTS}, and the baselines under different parameter settings.

\subsubsection{Default}
\begin{figure}[!htbp]
    \centering
    \subfigure[Time vs. Dataset]{
        \begin{minipage}[t]{0.45\columnwidth}
            \centering
            \vspace*{-7pt}
            \includegraphics[width=\columnwidth]{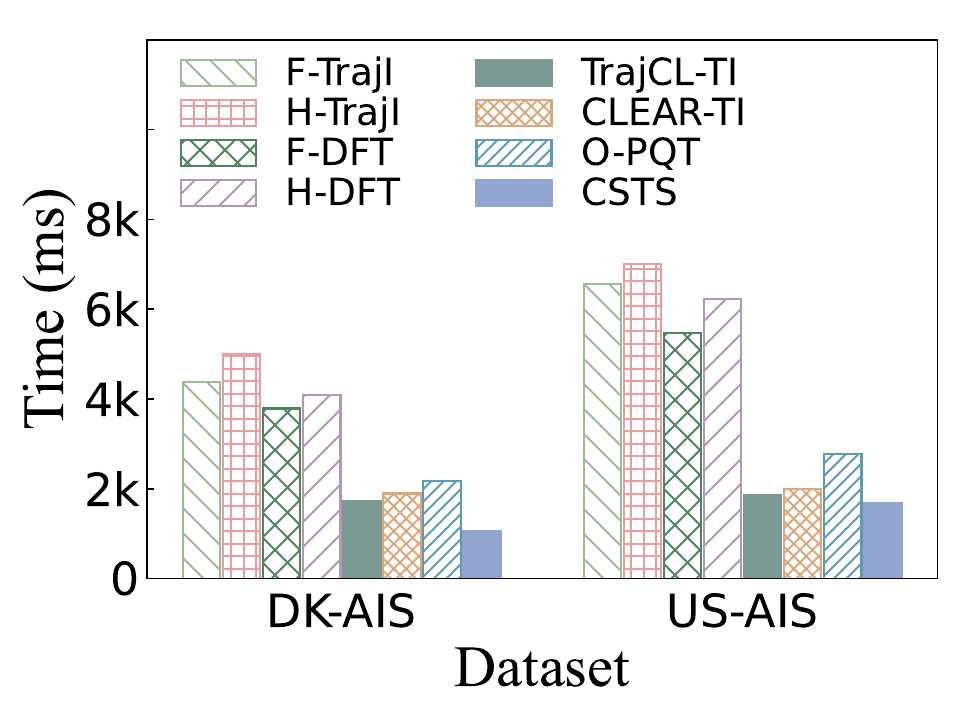}
            \label{fig:query_main_time}
            \vspace*{-3pt}
        \end{minipage}
    }
    \subfigure[Hit vs. Dataset]{
        \begin{minipage}[t]{0.45\columnwidth}
            \centering
            \vspace*{-7pt}
            \includegraphics[width=\columnwidth]{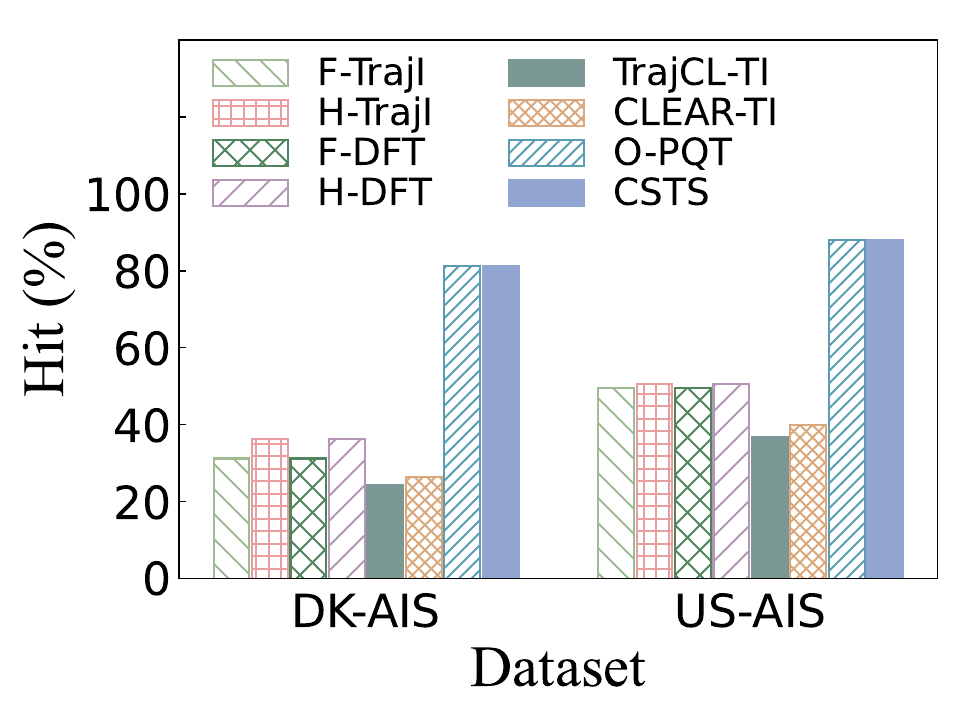}
            \label{fig:query_main_hit}
            \vspace*{-3pt}
        \end{minipage}
    }
    \centering
    \caption{Comparison of different methods on querying.}
    \label{fig:query_main}
    \vspace{-2mm}
\end{figure}

Fig.~\ref{fig:query_main_time} presents the query time for all methods on the two datasets under the default settings. \texttt{CSTS} demonstrates the best performance in terms of query time, attributed to its well-designed search algorithm with speed-up strategies. 
\texttt{O-PRT} has the longest search time, approximately 57 seconds on \texttt{DK-AIS} and 13 seconds on \texttt{US-AIS}, and its results are not displayed in the figure due to the significant disparity. 
\texttt{H-TrajI} and \texttt{F-TrajI}, which utilize an index based on entire trajectories, incur higher costs during the candidate search phase within the specified range. 
\texttt{H-DFT} and \texttt{F-DFT} perform relatively better than \texttt{H/F-TrajI} because the segment-based index offers increased efficiency during candidate searching. 
Both \texttt{H/F-TrajI} and \texttt{H/F-DFT} employ the Hausdorff or Fr{\'e}chet distance measures, which increase the search time during distance computations considerably, as they consider all point pairs within the trajectories. 
\texttt{TrajCL-TI} and \texttt{CLEAR-TI}, the two AI-based methods, exhibit strong efficiency since they leverage GPU capabilities, enabling faster distance computations. \texttt{O-PQT} takes longer than \texttt{CSTS}, indicating that the \texttt{SVTI} in \texttt{CSTS} is more effective than the Quad-tree based index.

Fig.~\ref{fig:query_main_hit} shows the hit rates of all methods on the two datasets under default settings. The hit rate of \texttt{CSTS} reaches 81\% on \texttt{DK-AIS} and 87\% on \texttt{US-AIS}, the highest among all methods. This superior performance is attributed to the proposed \texttt{OTRD} measure, which incorporates future movement trends of trajectories. 
\texttt{O-PQT} and \texttt{O-PRT} exhibit hit rates identical to that of \texttt{CSTS}, as they also employ the \texttt{OTRD} measure, which determines effectiveness in similar trajectory search. \texttt{H-TrajI} and \texttt{H-DFT} achieve the same hit rate because both use Hausdorff distance, while \texttt{F-TrajI} and \texttt{F-DFT} also perform equivalently in terms of hit rate due to their use of Fr{\'e}chet distance.
The Hausdorff and Fr{\'e}chet measures show comparable hit rates but perform worse than \texttt{OTRD} because they rely on static or full-path comparisons, making them a poor fit for evolving trajectories. The two AI-based similarity computation methods, \texttt{TrajCL-TI} and \texttt{CLEAR-TI}, achieve the lowest hit rates because they are designed to cluster trajectory variants and are not optimized for computing similarity across all original trajectories.

\begin{figure}[!htbp]
    \centering
    \subfigure[Time vs. $l_q$ (DK-AIS)]{
        \begin{minipage}[t]{0.45\columnwidth}
            \centering
            \vspace*{-7pt}
            \includegraphics[width=\columnwidth]{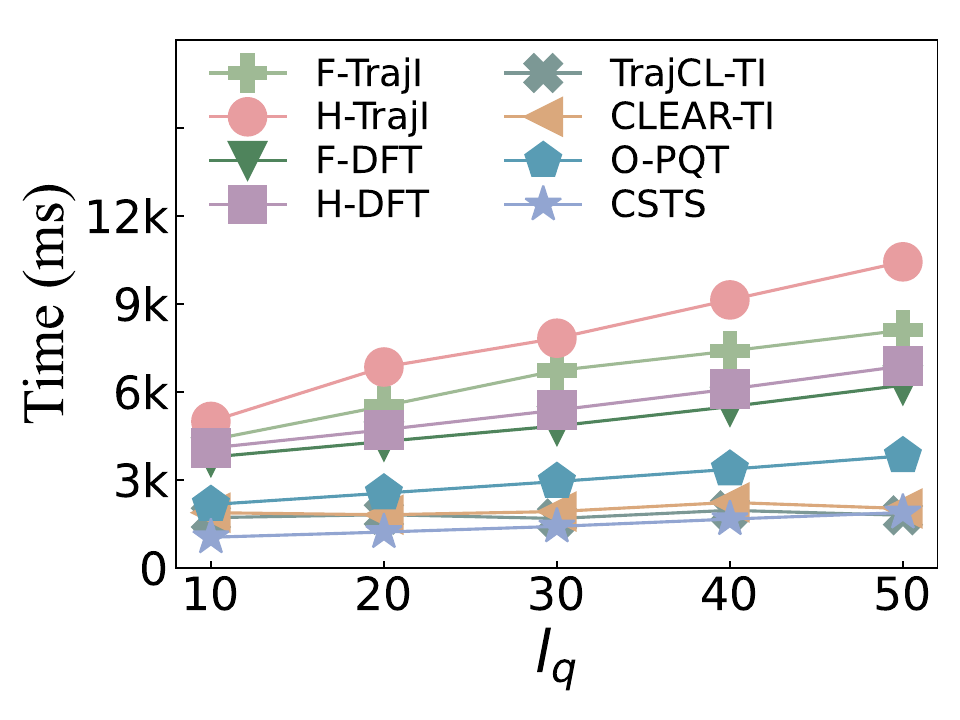}
            \label{fig:query_lq_time_DK}
            \vspace*{-3pt}
        \end{minipage}
    }
    \subfigure[Hit vs. $l_q$ (DK-AIS)]{
        \begin{minipage}[t]{0.45\columnwidth}
            \centering
            \vspace*{-7pt}
            \includegraphics[width=\columnwidth]{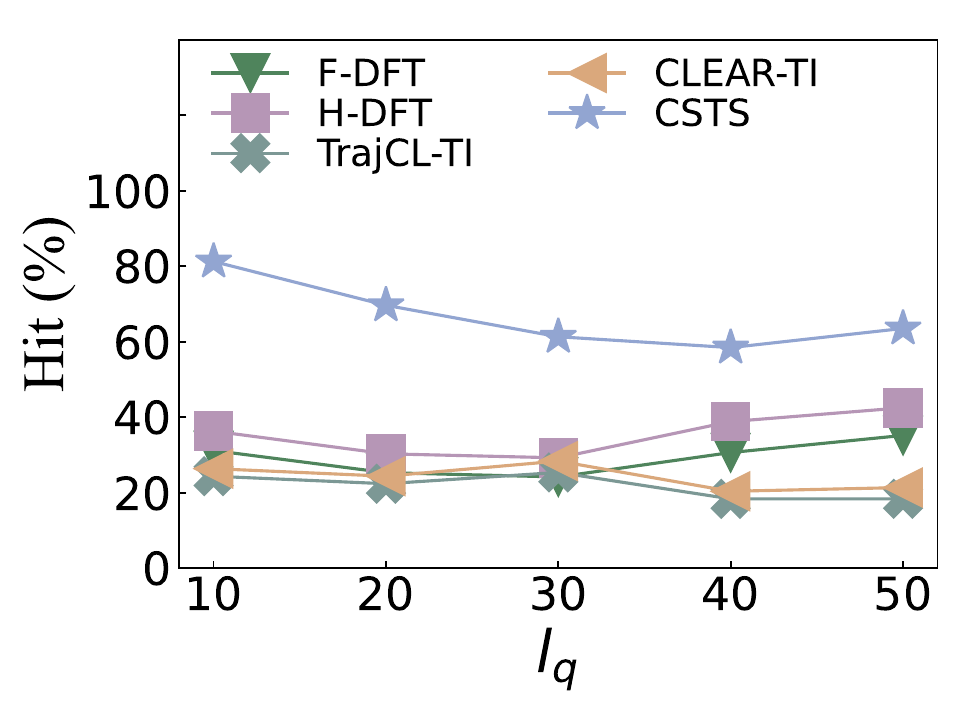}
            \label{fig:query_lq_hit_DK}
            \vspace*{-3pt}
        \end{minipage}
    }
    
    \subfigure[Time vs. $l_q$ (US-AIS)]{
        \begin{minipage}[t]{0.45\columnwidth}
            \centering
            \vspace*{-7pt}
            \includegraphics[width=\columnwidth]{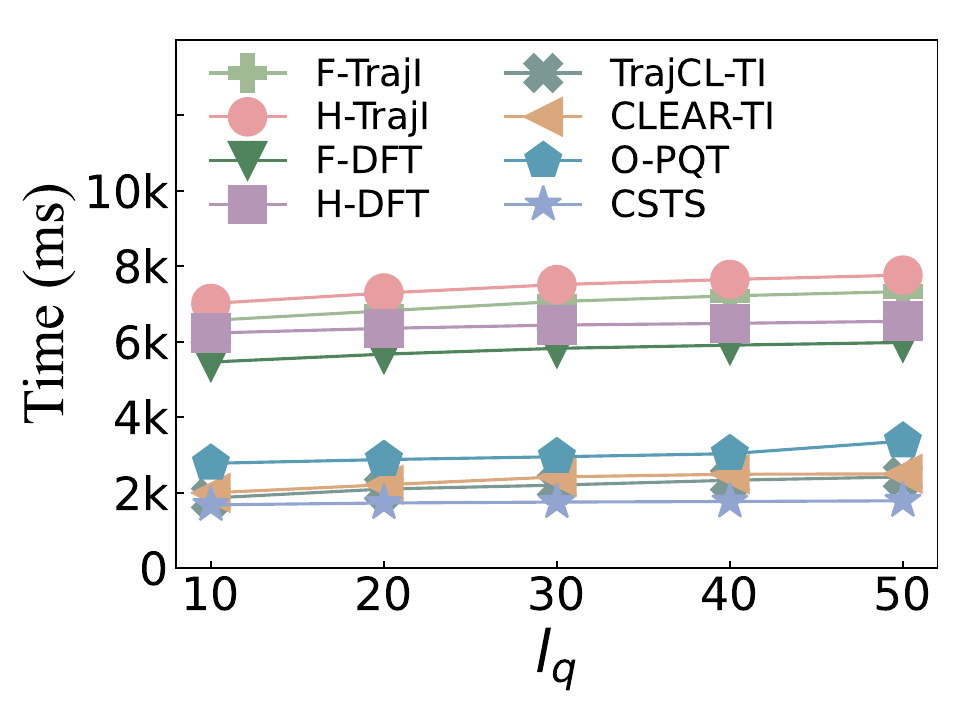}
            \label{fig:query_lq_time_US}
            \vspace*{-3pt}
        \end{minipage}
    }
    \subfigure[Hit vs. $l_q$ (US-AIS)]{
        \begin{minipage}[t]{0.45\columnwidth}
            \centering
            \vspace*{-7pt}
            \includegraphics[width=\columnwidth]{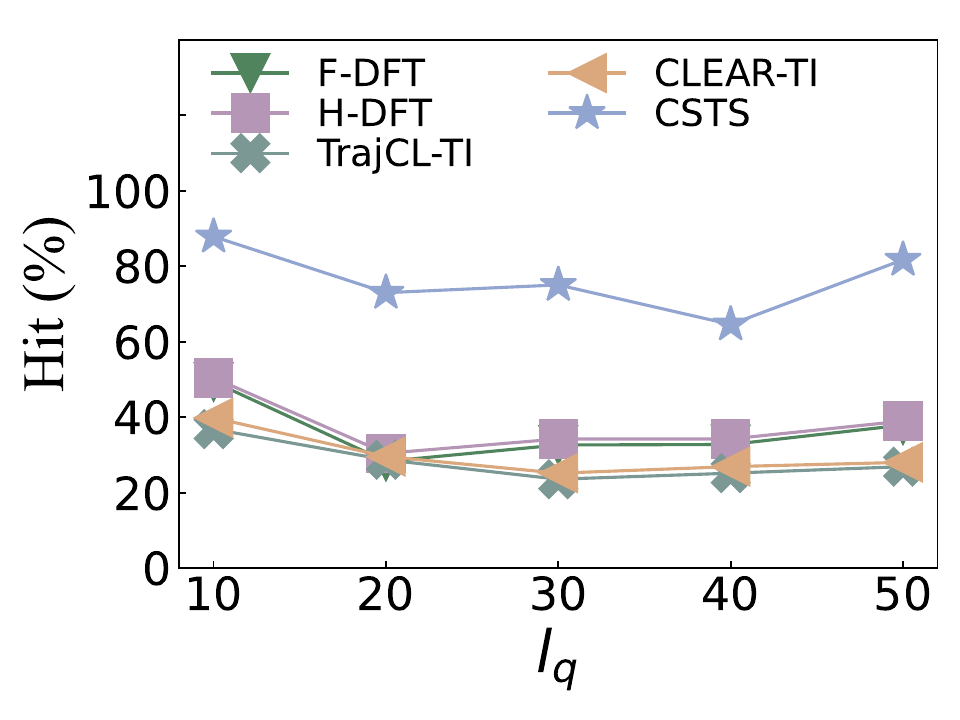}
            \label{fig:query_lq_hit_US}
            \vspace*{-3pt}
        \end{minipage}
    }
    \centering
    \caption{Effect of $l_q$.}
    \label{fig:query_lq}
\end{figure}

\subsubsection{Effect of $l_q$}
As shown in Figs.~\ref{fig:query_lq_time_DK} and~\ref{fig:query_lq_time_US}, the query times for all methods increase as the query length $l_q$ increases. Among the traditional methods, \texttt{CSTS} exhibits the smallest rate of increase because it only evaluates the historical portion of trajectories, avoiding point-by-point checks for the entire trajectory. The query time for \texttt{O-PQT} also increases slightly since it uses the same similarity measure as \texttt{CSTS}, but its reliance on a point-based quad-tree yields higher search costs.
\texttt{H/F-TrajI} and \texttt{H/F-DFT} exhibit a more pronounced growth in query time as $l_q$ increases, as each point in the historical part of the query trajectory is compared against the entire set of historical trajectories. In contrast, \texttt{TrajCL-TI} and \texttt{CLEAR-TI} show almost no variation in query time varying $l_q$, as these methods transform trajectories into fixed-length vectors, making them independent of query length.
Figs.~\ref{fig:query_lq_hit_DK} and~\ref{fig:query_lq_hit_US} reveal no discernible trend in the hit rate, suggesting that the effect of $l_q$ on hit rate is relatively minor.

\begin{figure}[!htbp]
    \centering
    \subfigure[Time vs. $r$ (DK-AIS)]{
        \begin{minipage}[t]{0.45\columnwidth}
            \centering
            \vspace*{-7pt}
            \includegraphics[width=\columnwidth]{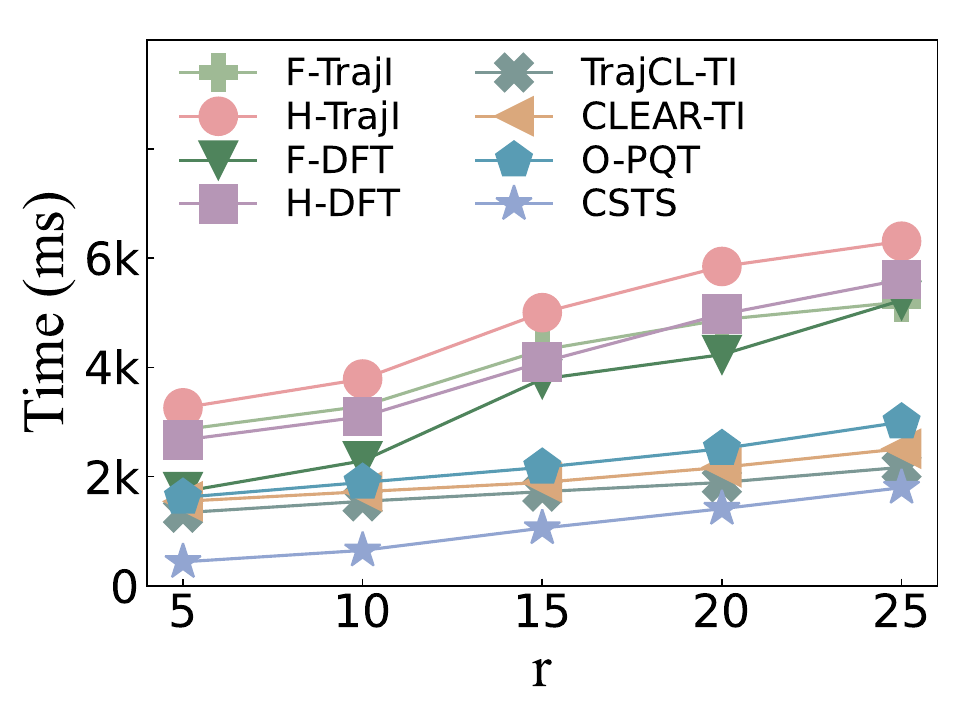}
            \label{fig:query_r_time_DK}
            \vspace*{-3pt}
        \end{minipage}
    }
    \subfigure[Hit vs. $r$ (DK-AIS)]{
        \begin{minipage}[t]{0.45\columnwidth}
            \centering
            \vspace*{-7pt}
            \includegraphics[width=\columnwidth]{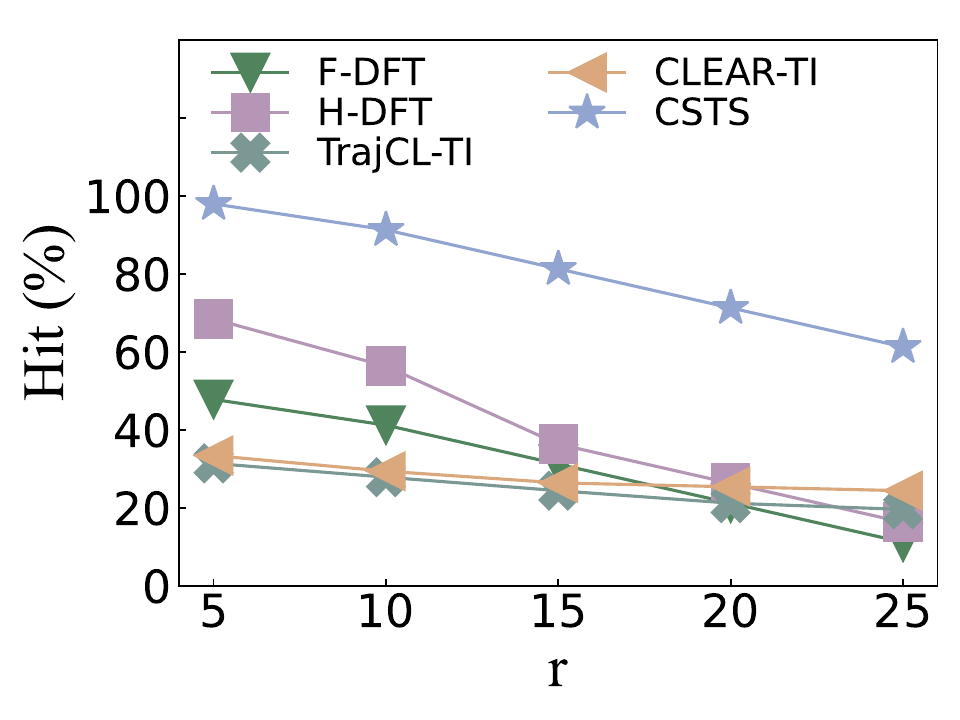}
            \label{fig:query_r_hit_DK}
            \vspace*{-3pt}
        \end{minipage}
    }

    \subfigure[Time vs. $r$ (US-AIS)]{
        \begin{minipage}[t]{0.45\columnwidth}
            \centering
            \vspace*{-7pt}
            \includegraphics[width=\columnwidth]{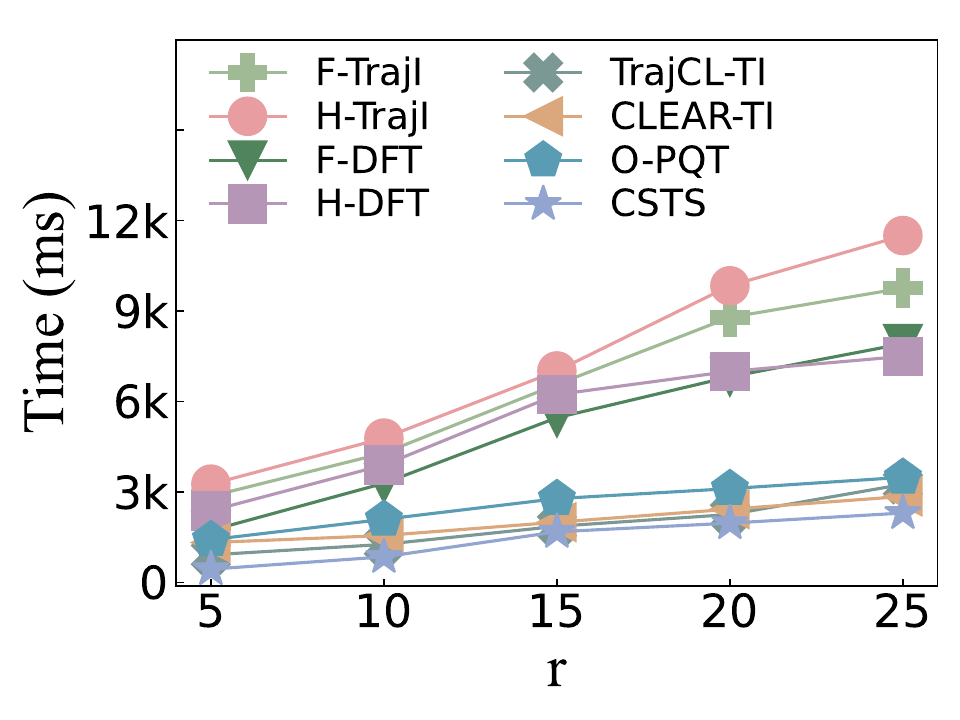}
            \label{fig:query_r_time_US}
            \vspace*{-3pt}
        \end{minipage}
    }
    \subfigure[Hit vs. $r$ (US-AIS)]{
        \begin{minipage}[t]{0.45\columnwidth}
            \centering
            \vspace*{-7pt}
            \includegraphics[width=\columnwidth]{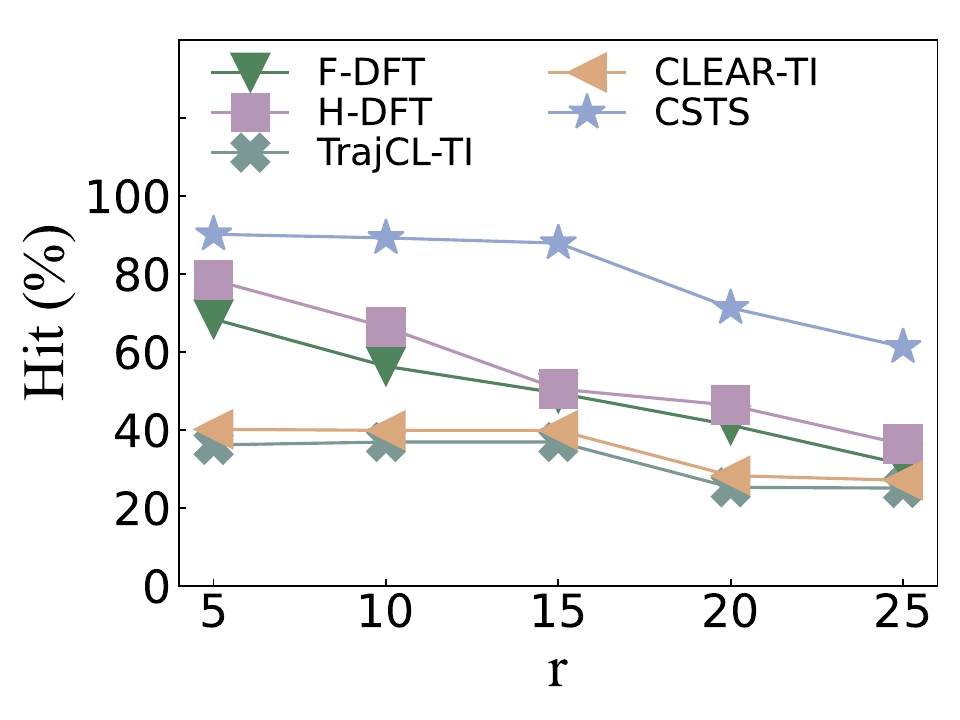}
            \label{fig:query_r_hit_US}
            \vspace*{-3pt}
        \end{minipage}
    }
    \centering
    \caption{Effect of $r$.}
    \label{fig:query_r}
    \vspace{-2mm}
\end{figure}

\subsubsection{Effect of $r$}
As shown in Figs.~\ref{fig:query_r_time_DK} and~\ref{fig:query_r_time_US}, the query time for all methods increases with the range $r$, as a larger range results in more candidates to evaluate. \texttt{H/F-TrajI} and \texttt{H/F-DFT} experience the fastest growth in query time due to the higher computational cost of similarity computations for each candidate compared to \texttt{OTRD} and the AI-based methods. Among all methods, \texttt{CSTS} remains the most efficient, even as the range increases, owing to its optimized search algorithm.
Figs.~\ref{fig:query_r_hit_DK} and~\ref{fig:query_r_hit_US} show that the hit rate decreases as the range increases. This is because a larger range includes more candidates. 
Some of these candidates might appear in the final similar trajectory results but not in the $k$NN of $o$'s next position, leading to a reduced hit rate.

\begin{figure}[!htbp]
    \centering
    \subfigure[Time vs. $k$ (DK-AIS)]{
        \begin{minipage}[t]{0.45\columnwidth}
            \centering
            \vspace*{-7pt}
            \includegraphics[width=1\columnwidth]{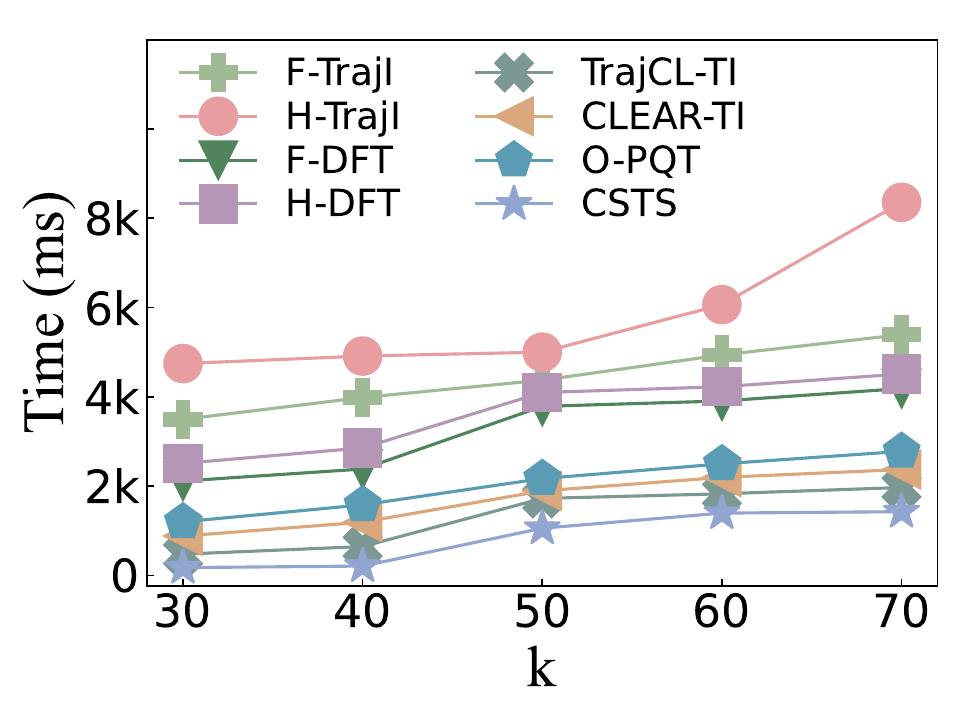}
            \label{fig:query_k_time_DK}
            \vspace*{-3pt}
        \end{minipage}
    }
    \subfigure[Hit vs. $k$ (DK-AIS)]{
        \begin{minipage}[t]{0.45\columnwidth}
            \centering
            \vspace*{-7pt}
            \includegraphics[width=1\columnwidth]{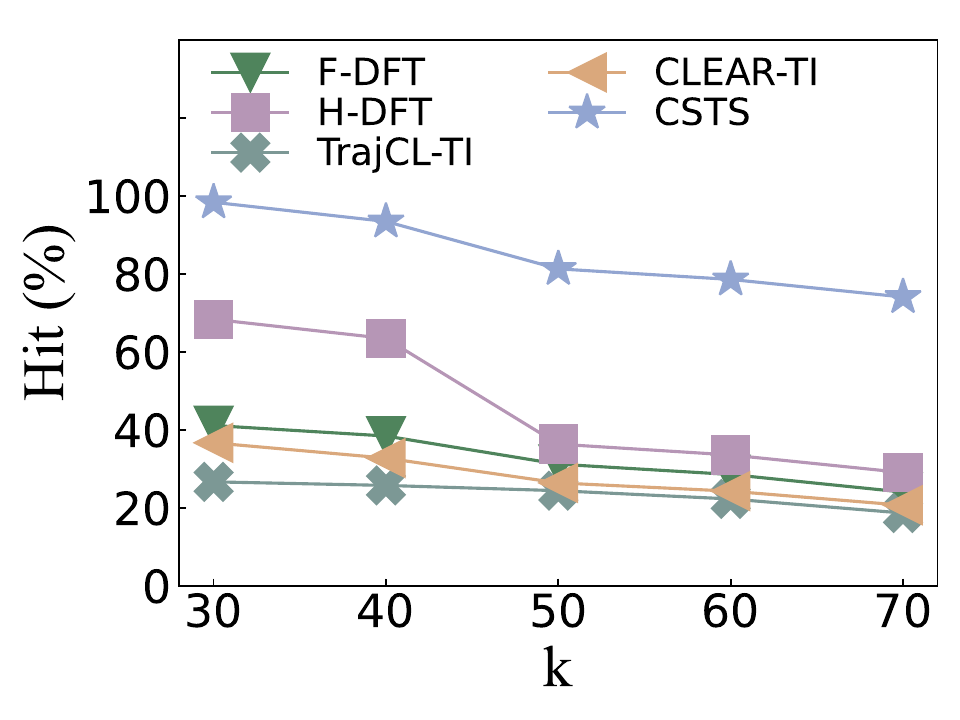}
            \label{fig:query_k_hit_DK}
            \vspace*{-3pt}
        \end{minipage}
    }

    \subfigure[Time vs. $k$ (US-AIS)]{
        \begin{minipage}[t]{0.45\columnwidth}
            \centering
            \vspace*{-7pt}
            \includegraphics[width=1\columnwidth]{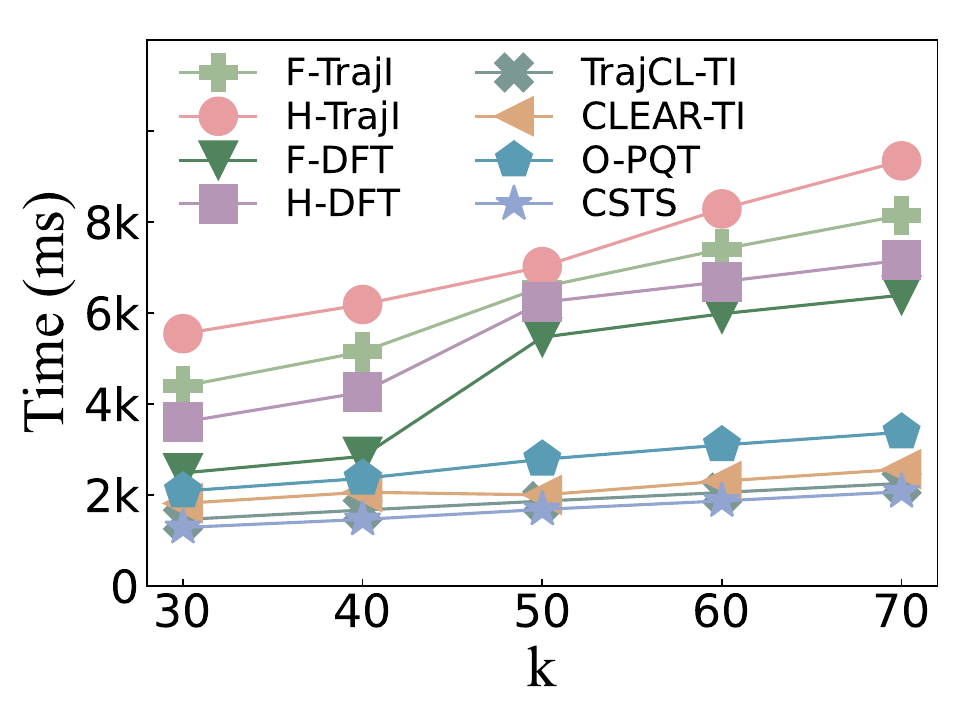}
            \label{fig:query_k_time_US}
            \vspace*{-3pt}
        \end{minipage}
    }
    \subfigure[Hit vs. $k$ (US-AIS)]{
        \begin{minipage}[t]{0.45\columnwidth}
            \centering
            \vspace*{-7pt}
            \includegraphics[width=1\columnwidth]{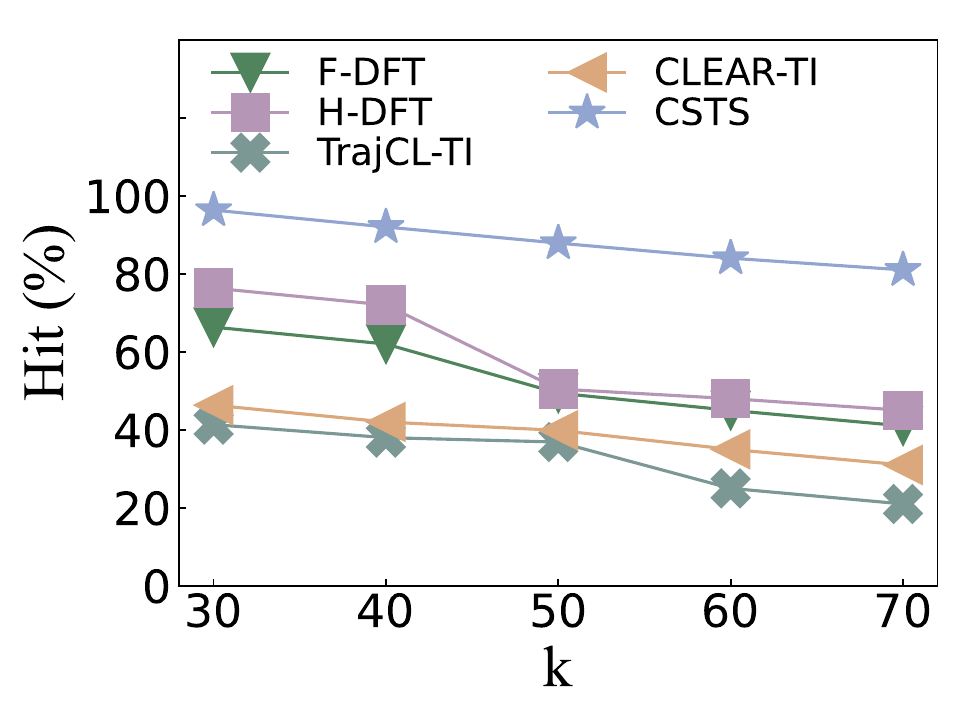}
            \label{fig:query_k_hit_US}
            \vspace*{-3pt}
        \end{minipage}
    }
    \centering
    \caption{Effect of $k$.}
    \label{fig:query_k}
    \vspace{-1mm}
\end{figure}

\subsubsection{Effect of $k$}
As shown in Fig.~\ref{fig:query_k}, increasing the value of $k$ leads to trends in query time and hit rate similar to those observed when increasing the value of $r$. Specifically, the query time increases, while the hit rate decreases. This occurs because the number of candidates is directly proportional to $k$, and a larger $k$ results in more candidates to evaluate, thereby increasing the time required for similarity computation. Moreover, the hit rate decreases as $k$ grows since $k$ serves as the denominator in the hit rate computation, making it more likely to produce a lower hit rate with larger $k$.

\subsubsection{Effect of Data Size}
\begin{figure}[!htbp]
    \centering
    \subfigure[Time vs. $\Lambda$ (US-AIS)]{
        \begin{minipage}[t]{0.45\columnwidth}
            \centering
            \vspace*{-7pt}
            \includegraphics[width=\columnwidth]{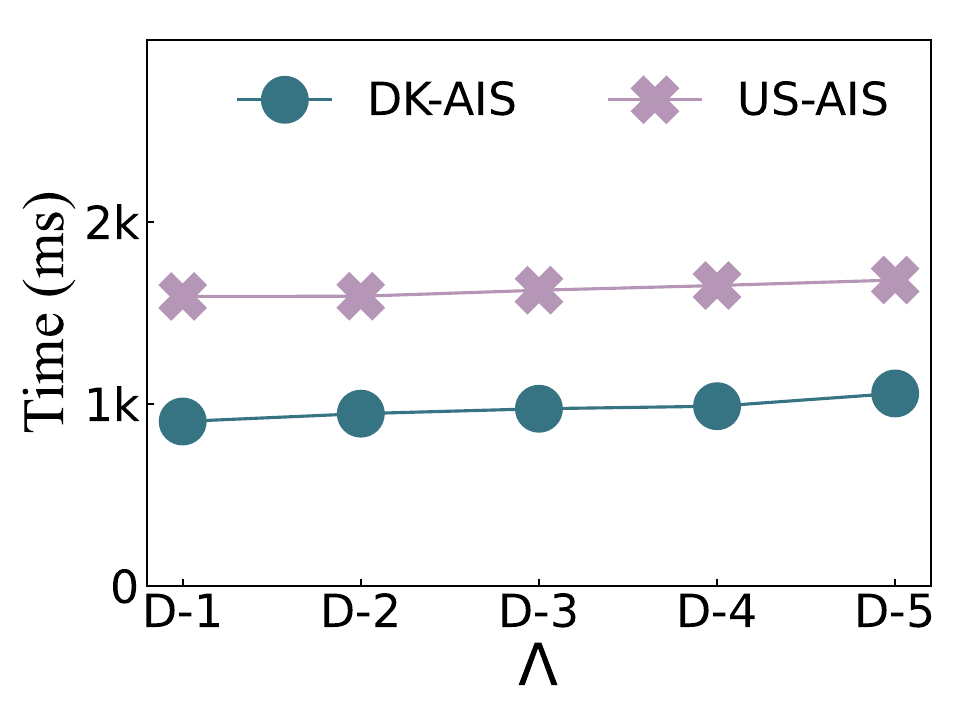}
            \label{fig:query_scalability_time}
            \vspace*{-3pt}
        \end{minipage}
    }
    \subfigure[Hit vs. $\Lambda$ (US-AIS)]{
        \begin{minipage}[t]{0.45\columnwidth}
            \centering
            \vspace*{-7pt}
            \includegraphics[width=\columnwidth]{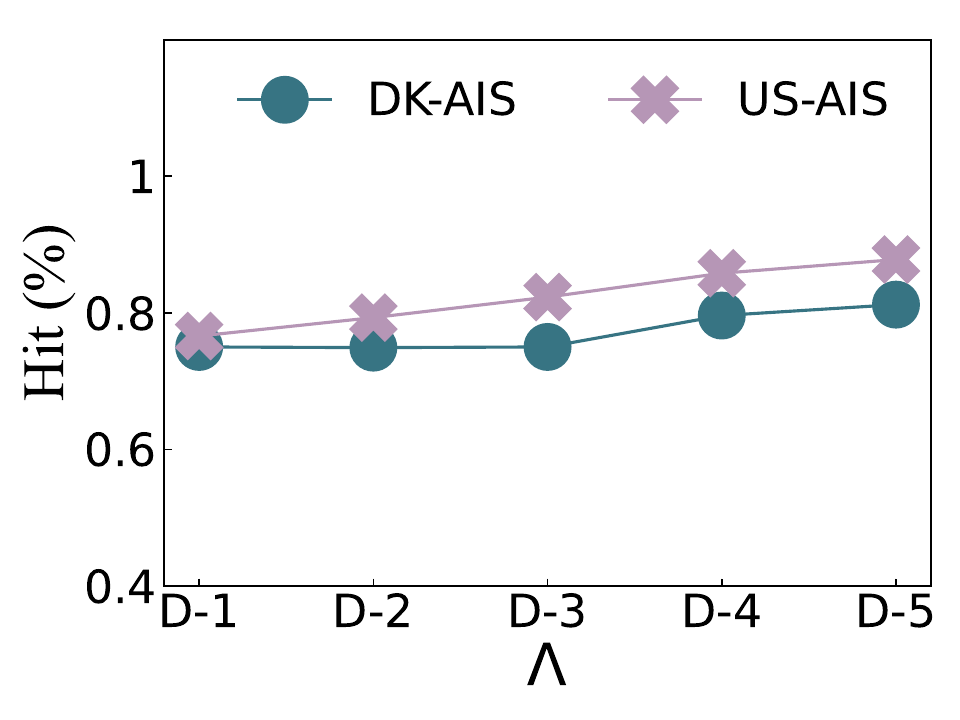}
            \label{fig:query_scalability_hit}
            \vspace*{-3pt}
        \end{minipage}
    }
    \centering
    \caption{Effect of data size.}
    \label{fig:query_scalability}
    \vspace{-2mm}
\end{figure}

We evaluate \texttt{CSTS} under varying data sizes, as shown in Fig.~\ref{fig:query_scalability}. The settings for datasets D-1 through D-5 remain consistent with those described in Section~\ref{sss:datasize}. As the data size increases, both the query time and the hit rate remain relatively stable. This is because the number of candidates remains consistent regardless of the dataset size. Consequently, the query time is primarily determined by the similarity computation rather than the candidate-finding phase, which incurs negligible overhead.
\subsubsection{Effect of $\theta$ and $\alpha$}
\begin{figure}[!htbp]
    \centering
    \subfigure[Hit vs. $\theta$]{
        \begin{minipage}[t]{0.45\columnwidth}
            \centering
            \vspace*{-7pt}
            \includegraphics[width=\columnwidth]{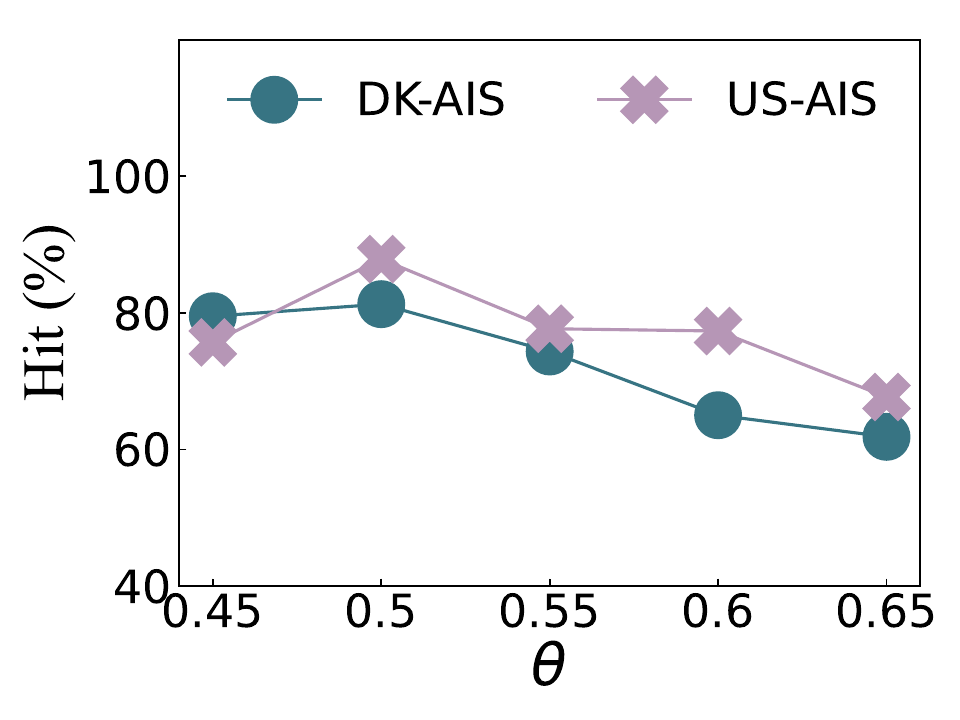}
            \label{fig:query_theta_hit}
            \vspace*{-3pt}
        \end{minipage}
    }
    \subfigure[Hit vs. $\alpha$]{
        \begin{minipage}[t]{0.45\columnwidth}
            \centering
            \vspace*{-7pt}
            \includegraphics[width=\columnwidth]{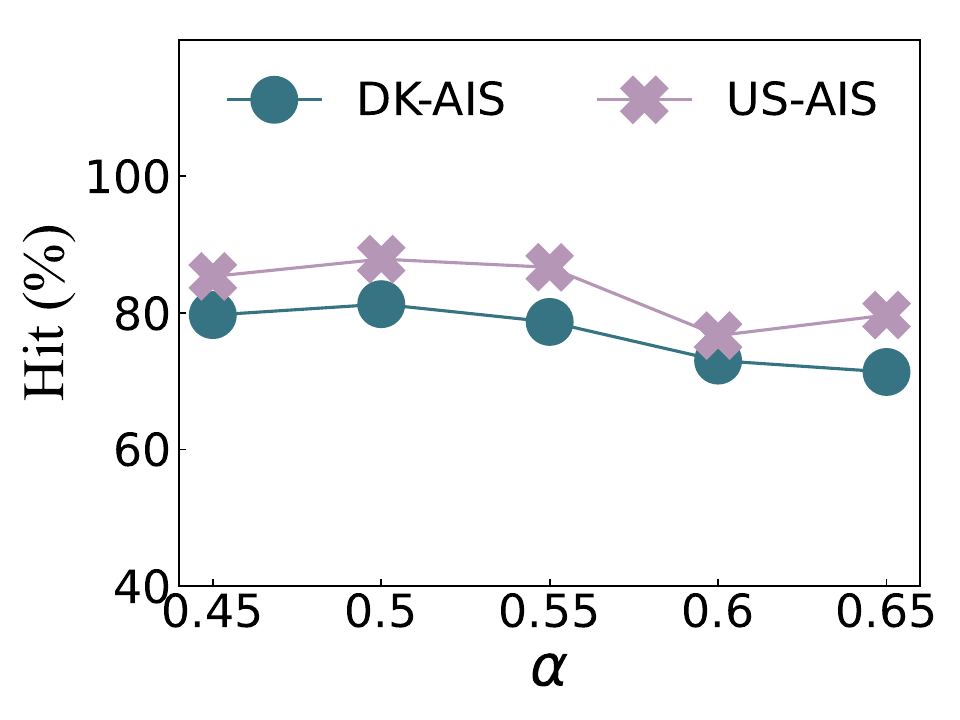}
            \label{fig:query_alpha_hit}
            \vspace*{-3pt}
        \end{minipage}
    }
    \centering
    \caption{Effect of $\theta$ and $\alpha$.}
    \label{fig:query_theta_alpha}
    \vspace{-5mm}
\end{figure}

We examine the impact of the decay factor $\theta$ and the trade-off factor $\alpha$ on \texttt{CSTS}. Since these parameters do not influence efficiency, only their effects on the hit rate are reported.
As shown in Fig.~\ref{fig:query_theta_hit}, the hit rate decreases slightly as $\theta$ increases for both datasets. This is because a smaller $\theta$ results in a higher decay for earlier points in the historical portion of the query object. Consequently, points closer to the current position of the query object are given higher weight during similarity computation, leading to more accurate trajectory matches.

Similarly, Fig.~\ref{fig:query_alpha_hit} shows a slight decline in the hit rate as $\alpha$ increases. This indicates that the Historical Trajectory Distance plays a more significant role in the similarity measure compared to the coarser-grained Target-Trajectory Distance.

\begin{figure}[h]
    \centering
    \includegraphics[width=0.9\columnwidth]{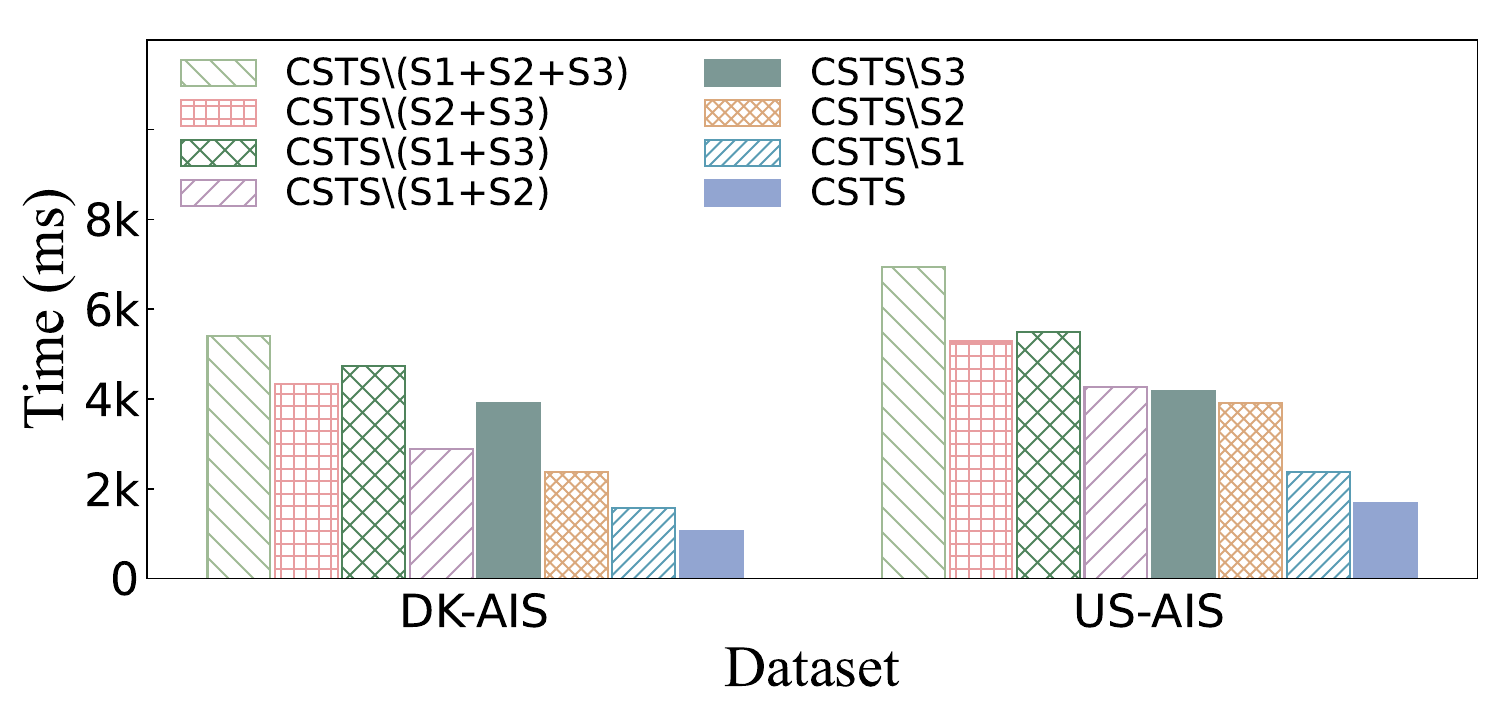}
    \vspace*{-7pt}
    \caption{Effect of speed-up strategies.}
    \label{fig:query_ablation_time}
    \vspace{-5mm}
\end{figure}
\subsection{Ablation Study}
\label{ssec:ablatin_study}
By default, \texttt{CSTS} incorporates segment pruning (Strategy~\ref{Strategy: S1}), $k$-bound pruning (Strategy~\ref{Strategy: S2}), and the incremental rule (Strategy~\ref{Strategy: S3}). These three strategies are solely designed to improve efficiency and do not reduce the hit rate. 
In Fig.~\ref{fig:query_ablation_time}, we report their impact on query time. 
Among the strategies, the incremental rule (S3) proves to be the most effective, reducing the query time by at least threefold. This is the evidence of the efficiency of the incremental rule in optimizing \texttt{CSTS}. The second-most effective strategy is $k$-bound pruning (S2), which shows comparable performance to S3. Segment pruning (S1), however, provides relatively less improvement on both datasets. This is because it introduces additional computational costs to compute the distance between a point and the minimum bounding rectangle of a segment.

\begin{figure}[h]
    \centering
    \subfigure[Time vs. $g$]{
        \begin{minipage}[t]{0.45\columnwidth}
            \centering
            \vspace*{-7pt}
            \includegraphics[width=\columnwidth]{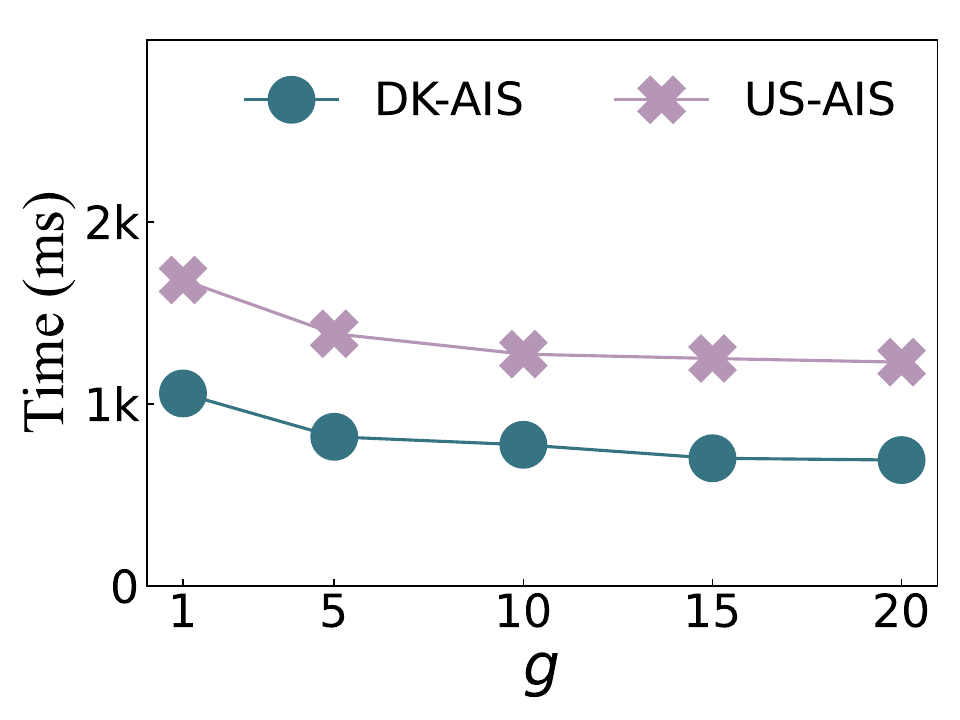}
            \label{fig:query_granularity_time}
            \vspace*{-3pt}
        \end{minipage}
    }
    \subfigure[Hit vs. $g$]{
        \begin{minipage}[t]{0.45\columnwidth}
            \centering
            \vspace*{-7pt}
            \includegraphics[width=\columnwidth]{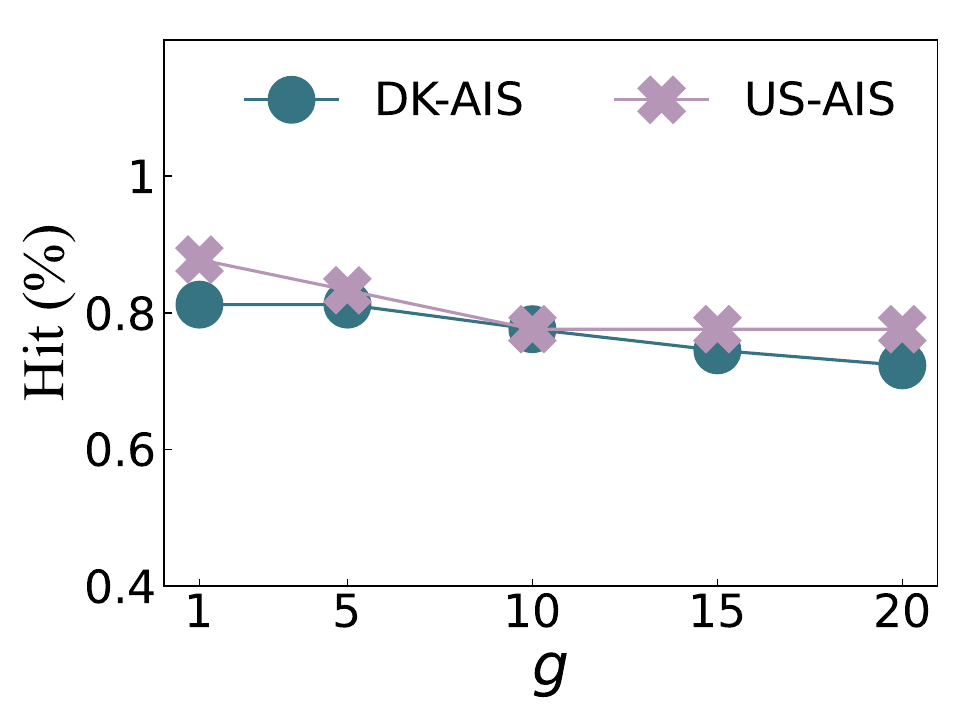}
            \label{fig:query_granularity_hit}
            \vspace*{-3pt}
        \end{minipage}
    }
    \centering
    \caption{Effect of granularity.}
    \label{fig:query_granularity}
    \vspace{-2mm}
\end{figure}

The granularity adjustment (Strategy~\ref{Strategy: S4}) is not applied in \texttt{CSTS} by default, as it may introduce errors. To evaluate the effectiveness of this strategy, we vary the granularity parameter $g$. Referring to Fig.~\ref{fig:query_granularity}, both query time and hit rate decrease as $g$ increases. With a larger $g$, more points are disregarded during similarity computation, resulting in reduced query time. However, the hit rate declines because ignoring points introduces errors.

\begin{figure}[h]
    \centering
    \includegraphics[width=\columnwidth]{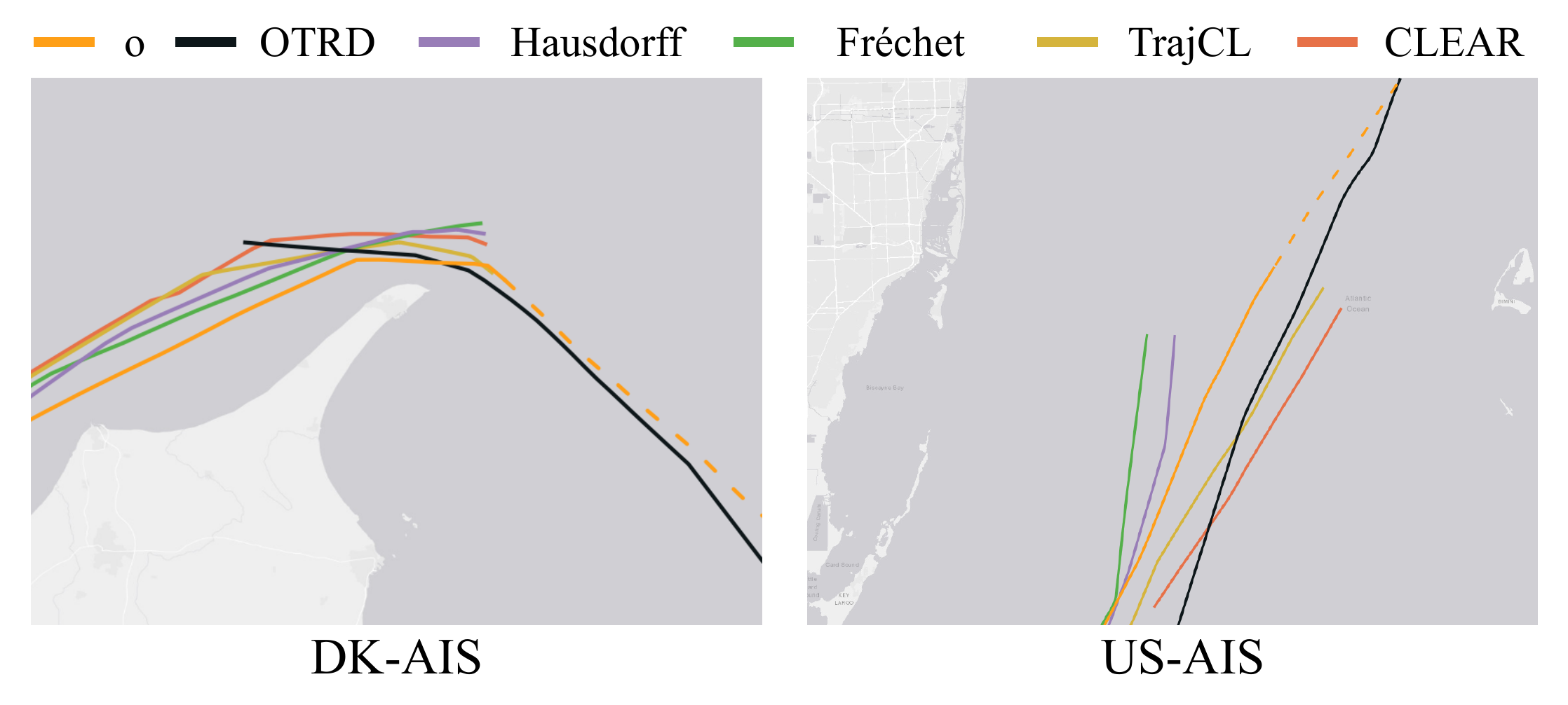}
    \caption{Case study.}
    \label{fig:case_study}
    \vspace{-5mm}
\end{figure}

\subsection{Case Study on Similarity Measures}
\label{ssec:case_study_realdata}
Fig.~\ref{fig:case_study} presents case studies on \texttt{DK-AIS} and \texttt{US-AIS}, illustrating the differences in results when applying different similarity measures. The orange line represents a moving object $o$, including both its current trajectory and the projected destination. The black line indicates the most similar trajectory identified using the \texttt{OTRD} measure.
In comparison with other similarity measures such as Hausdorff, Fr{\'e}chet, \texttt{TrajCL}, and \texttt{CLEAR}, the \texttt{OTRD} measure demonstrates a clear advantage. The other four measures focus predominantly on aligning with $o$’s current trajectory, resulting in matches that are less aligned with $o$’s overall path. Conversely, the trajectory identified by \texttt{OTRD} not only aligns closely with $o$’s current trajectory but also incorporates the future trend and destination, offering a more holistic similarity result.
This highlights the strength of \texttt{OTRD} at capturing the dynamic and predictive aspects of trajectory comparison, making it suitable for applications where future trends and endpoints are important.

\section{Related Work}
\label{sec:related}

\noindent\textbf{Trajectory similarity measures.}
Trajectory similarity measures form the foundation of trajectory analysis. 
Dynamic Time Warping~\cite{yi1998efficient, ying2016simple} allows for flexible alignment and comparison of time-series data, accommodating variations. 
Longest Common Subsequence~\cite{vlachos2002discovering} offers robustness to noise and spatiotemporal deformations by identifying the longest subsequence shared between trajectories.
Additionally, Fr{\'e}chet distance \cite{alt1995computing, de2013fast, agarwal2014computing} and 
edit distance-based measures \cite{chen2004marriage, chen2005robust, ranu2015indexing} provide nuanced assessments of trajectory similarity, considering geometric constraints and structural differences. 
The Hausdorff distance \cite{hangouet1995computation, bai2011polyline} measures how far the points of one set are from the points of the other set.
These measures \cite{toohey2015trajectory, magdy2015review} serve as the cornerstone for trajectory analysis.
However, these measures fail to consider the significance of individual points, the trends within historical trajectories, and the future movement patterns of objects.

\noindent\textbf{Learning-based trajectory similarity computation.}
Recent advancements in trajectory similarity search have explored learning-based approaches~\cite{li2018deep, fu2020trembr, liu2020representation, deng2022efficient, fang2022spatio, mao2022jointly, li2024clear} to improve the accuracy and efficiency of similarity computations. 
These methods often leverage deep learning models to learn representations of trajectories from raw spatiotemporal data. 
However, this method has several limitations that make it less suitable for real-time, continuous vessel trajectory search. First, these methods often rely on down-sampling or distortion of trajectories, which leads to the identification of trajectories as similar, even when they lack true alignment with the original paths. This compromises their ability to accurately capture the underlying similarity between the original trajectories. Second, these methods typically focus on computing the similarity between two given trajectories rather than provide efficiently similarity search method on a large trajectory database. Third, they are not well-suited for real-time processing. The continuous updates in vessel trajectories would necessitate frequent model re-training or recalibration, imposing substantial computational overhead and significantly degrading performance. Finally, learning-based models are generally designed for offline batch processing, making them unsuitable for dynamic and real-time maritime navigation.

\noindent\textbf{Similar trajectory search.}
Studies on similar trajectory search~\cite{ding2008efficient, xie2017distributed, wang2018torch, li2020vessel, he2022trass, luo2023vessel} and similarity join~\cite{bakalov2005efficient, bakalov2005time, ta2017signature, shang2017trajectory} have proposed various techniques to identify similar movement patterns. 
Some approaches~\cite{hwang2006searching, hwang2005spatio, xie2017distributed, yuan2019distributed, han2021graph, he2022trass, fang2022spatio} are widely used in domains like vehicle tracking and urban traffic monitoring, particularly in road networks, where vehicle movements are constrained by structured paths, making it easier to organize and index trajectories based on road segments, but not suitable for vessel trajectories. 
Furthermore, they mainly designed for existing trajectory similarity methods to compare entire trajectories, which is not suitable for real-time similar vessel trajectory search. 
These methods are less effective for real-time applications, where the trajectory evolves over time and must be analyzed incrementally. 
As a result, current trajectory similarity search methods are not well-suited to solve the problem of real-time, continuous vessel trajectory search.

\section{Conclusion and Future Work}
\label{sec:conclusion}

This paper presents \texttt{ACTIVE}, a robust framework for real-time continuous trajectory similarity search for vessels. By introducing the \texttt{OTRD} measure, \texttt{ACTIVE} focuses on predictive, forward-looking trajectory analysis, overcoming the limitations of traditional retrospective methods. We propose the \texttt{CSTS} algorithm for continuous search of similar historical trajectories. The developed index, \texttt{SVTI}, along with four speed-up strategies, enhances the system's efficiency and scalability, enabling rapid and accurate similarity computations. Experiments on real-world AIS datasets demonstrate that \texttt{ACTIVE} outperforms state-of-the-art methods by up to 70\% in terms of query time, 60\% in terms of hit rate, and reduces index construction costs significantly.

In future, it is of interest to apply the proposed continuous similarity search to real-time maritime applications, such as traffic monitoring, collision avoidance, and route optimization, enabling dynamic and proactive decision-making in large-scale vessel trajectory analysis.

\section*{Acknowledgments}
We acknowledge the support of European Union’s funded Project MobiSpaces under grant agreement no 101070279.



\bibliographystyle{IEEEtran}
\bibliography{ref}

\clearpage
\newpage
\appendix
\section{Evaluation}
\label{sec:appendixB}
\subsection{Evaluation on \texttt{SVTI}}
\label{ssec:evalutation_index}
\subsubsection{Settings}
We evaluate \texttt{SVTI} and four traditional indexing techniques—\texttt{PQT}, \texttt{PRT}, \texttt{TrajI}, and \texttt{DFT}—in terms of index construction time and index size. To assess the performance of \texttt{SVTI}, we vary three parameters with the following settings (default values in bold):
\begin{itemize}
    \item $l_\textit{min}$: The minimum segment lengths. Setting: 20, 25, \textbf{30}, 35, 40.
    \item $l_\textit{max}$: The maximum segment lengths. Setting: 40, 45, \textbf{50}, 55, 60.
    \item Data size ($\Lambda$): The number of the points in the AIS data. Five different data sizes are evaluated: D-1, D-2, D-3, D-4, and D-5, corresponding to approximately $20\%$, $40\%$, $60\%$, $80\%$, and $100\%$ of the original dataset.
\end{itemize}

\subsubsection{Default}

\begin{figure}[!htbp]
    \centering
    \subfigure[Time vs. Dataset]{
        \begin{minipage}[t]{0.45\columnwidth}
            \centering
            \includegraphics[width=\columnwidth]{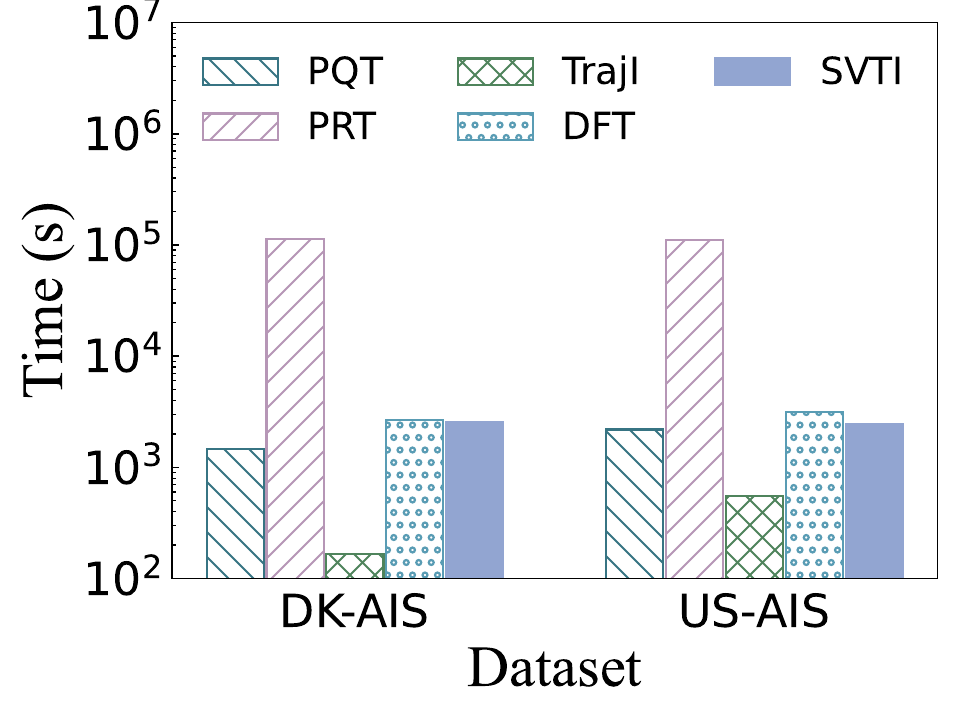}
            \label{fig:index_main_time}
        \end{minipage}
    }
    \subfigure[Size vs. Dataset]{
        \begin{minipage}[t]{0.45\columnwidth}
            \centering
            \includegraphics[width=\columnwidth]{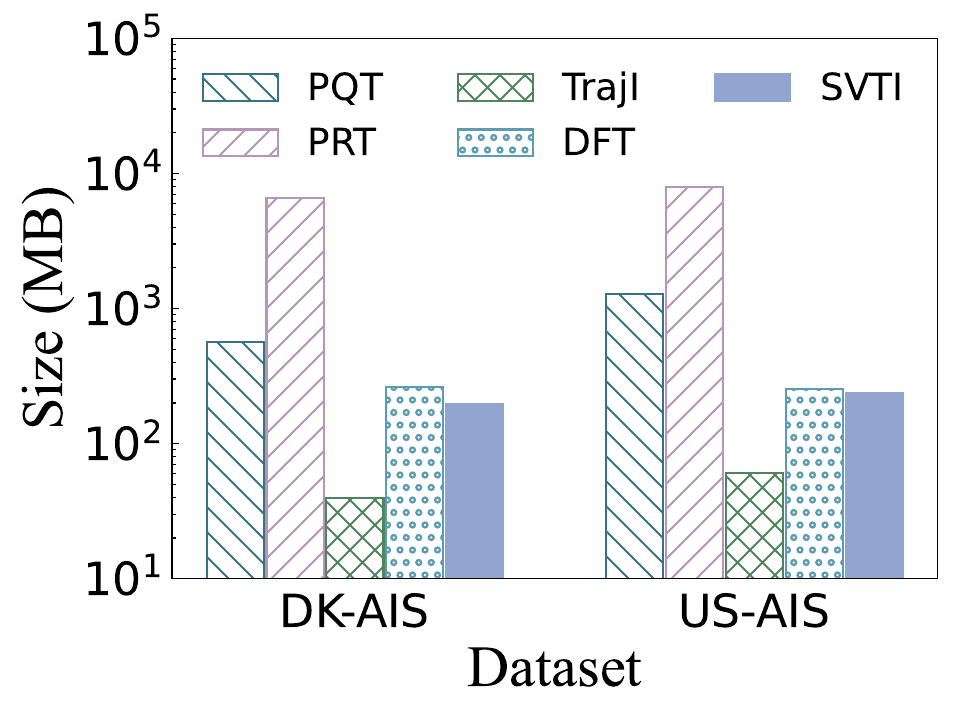}
            \label{fig:index_main_size}
        \end{minipage}
    }
    \centering
    \caption{Comparison of index construction time and size.}
    \label{fig:index_default}
\end{figure}

The performance of the different index methods under the default parameter settings is shown in Fig.~\ref{fig:index_default}. 
In terms of construction time, \texttt{SVTI} takes 42 minutes to build on the \texttt{DK-AIS} dataset and 40 minutes on the \texttt{US-AIS} dataset.
The index sizes of \texttt{SVTI} are 196 MB and 235 MB on \texttt{DK-AIS} and \texttt{US-AIS}, respectively. Among the five indices, \texttt{PRT} has the longest construction time, exceeding one day for both datasets due to its point-based design. Consequently, \texttt{PRT} also has the largest index size. 
Next, \texttt{TrajI} is the most efficient index to construct, as it is based on entire trajectories. This design minimizes insertion times, making its construction time directly proportional to the number of insertions. 
The original segmentation method for \texttt{DFT}, designed for road networks, cannot be applied directly to AIS data, so our segment partitioning method is used. 
Therefore, the construction time and index size of \texttt{DFT} are comparable to those of \texttt{SVTI}.
Lastly, \texttt{PQT} exhibits relatively higher efficiency compared to \texttt{PRT} because the Quad-tree based \texttt{PQT} benefits from faster theoretical construction times compared to the R-tree based \texttt{PRT}. 
\subsubsection{Effect of Data Size.}\label{sss:datasize}
\begin{figure}[!htbp]
    \centering
    \subfigure[Time vs. $\Lambda$]{
        \begin{minipage}[t]{0.45\columnwidth}
            \centering
            \includegraphics[width=\columnwidth]{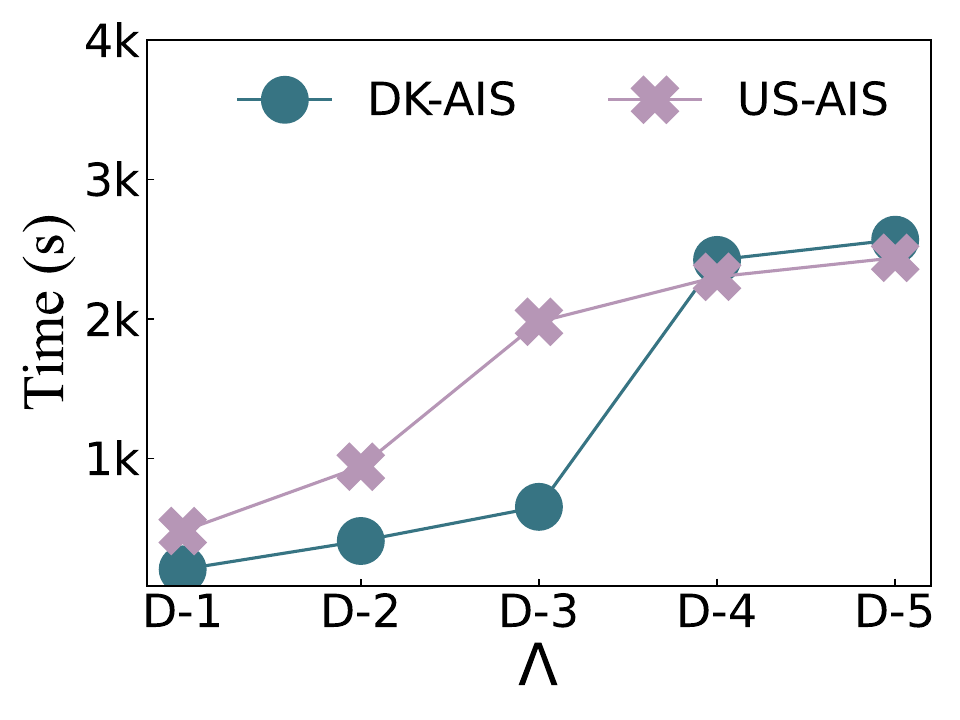}
            \label{fig:index_scalability_time}
        \end{minipage}
    }
    \subfigure[Size vs. $\Lambda$]{
        \begin{minipage}[t]{0.45\columnwidth}
            \centering
            \includegraphics[width=\columnwidth]{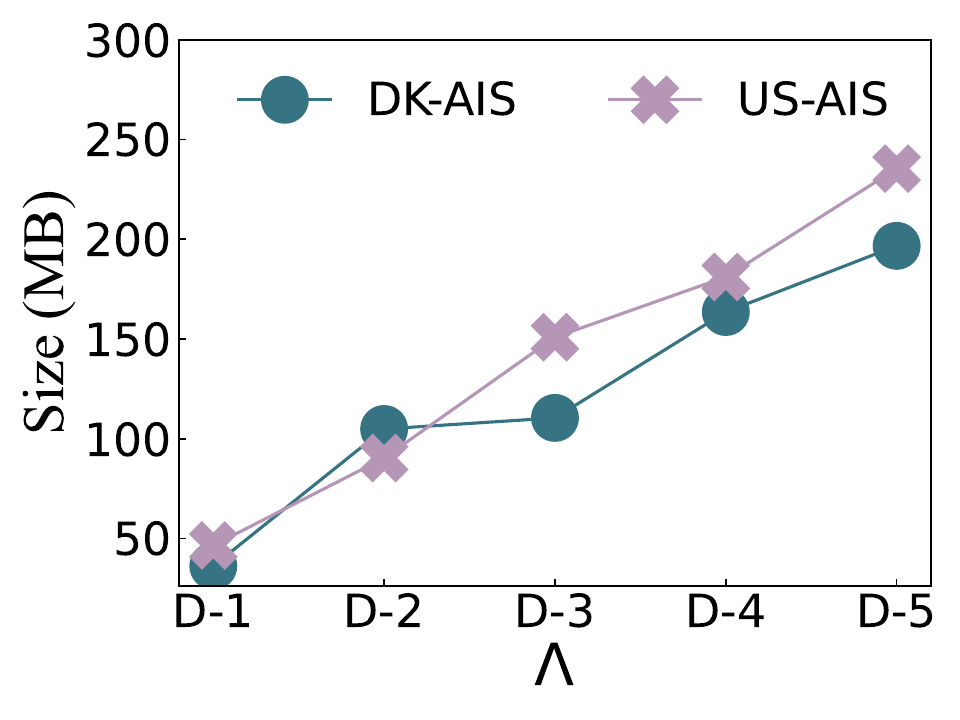}
            \label{fig:index_scalability_size}
        \end{minipage}
    }
    \centering
    \caption{Effect of data size.}
    \label{fig:index_scalability}
\end{figure}

We analyze the effect of varying data sizes on construction time and index size, with results shown in Figs.~\ref{fig:index_scalability_time} and~\ref{fig:index_scalability_size}, respectively.
The datasets D-1, D-2, D-3, D-4, and D-5 contain approximately 43, 86, 130, 172, and 216 million points for \texttt{DK-AIS}, and 31, 62, 93, 123, and 154 million points for \texttt{US-AIS}. As the data size increases, the number of segments and corresponding insertions also grow. This leads to a proportional increase in both construction time and index size.

\subsubsection{Effect of $l_\textit{max}$ and $l_\text{min}$}
\begin{figure}[!htbp]
    \centering
    \subfigure[Time vs. $l_\text{min}$]{
        \begin{minipage}[t]{0.45\columnwidth}
            \centering
            \includegraphics[width=\columnwidth]{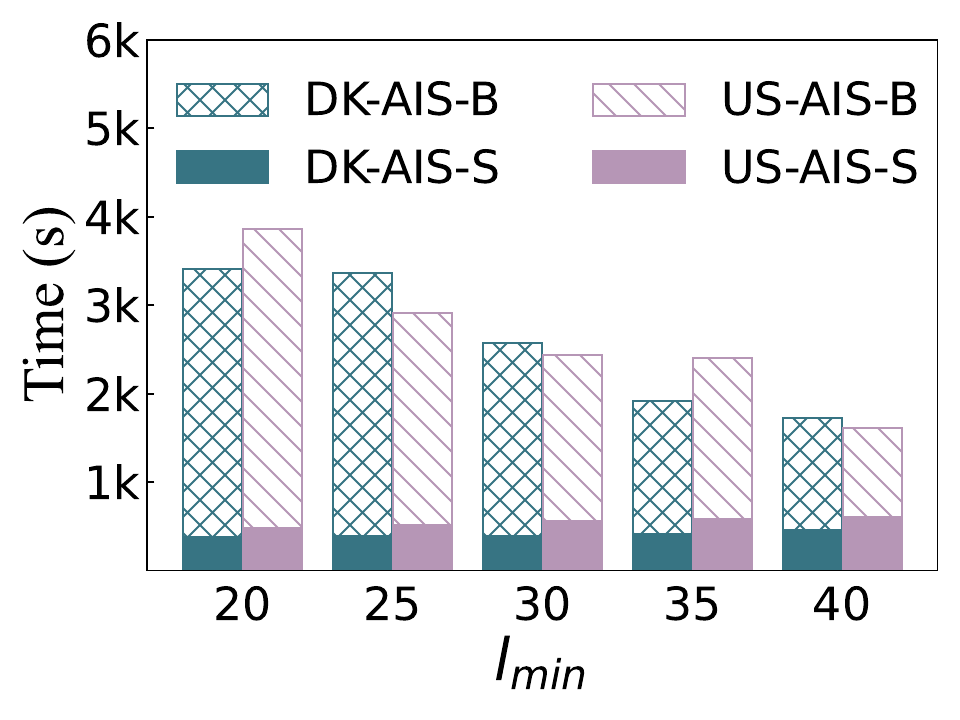}
            \label{fig:index_lm_time}
        \end{minipage}
    }
    \subfigure[Size vs. $l_\text{min}$]{
        \begin{minipage}[t]{0.45\columnwidth}
            \centering
            \includegraphics[width=\columnwidth]{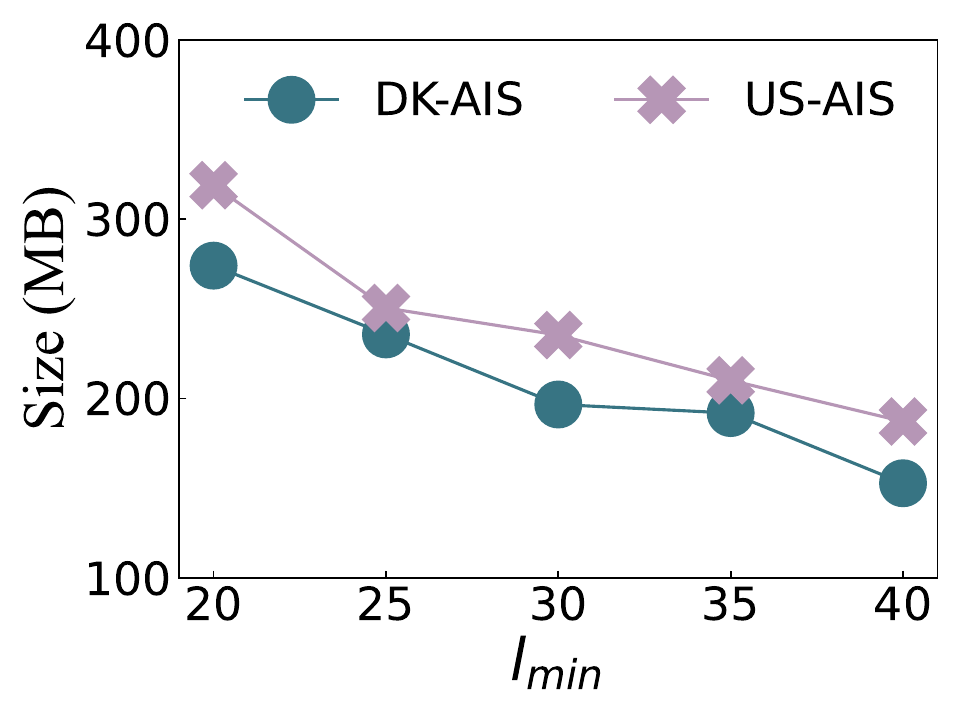}
            \label{fig:index_lm_size}
        \end{minipage}
    }
    \centering
    \caption{Effect of $l_\text{min}$ ($l_\textit{max} = l_\text{min} + 20$).}
    \label{fig:index_lm}
\end{figure}

\begin{figure}[!htbp]
    \centering
    \subfigure[Time vs. $l_\textit{max}$]{
        \begin{minipage}[t]{0.45\columnwidth}
            \centering
            \includegraphics[width=\columnwidth]{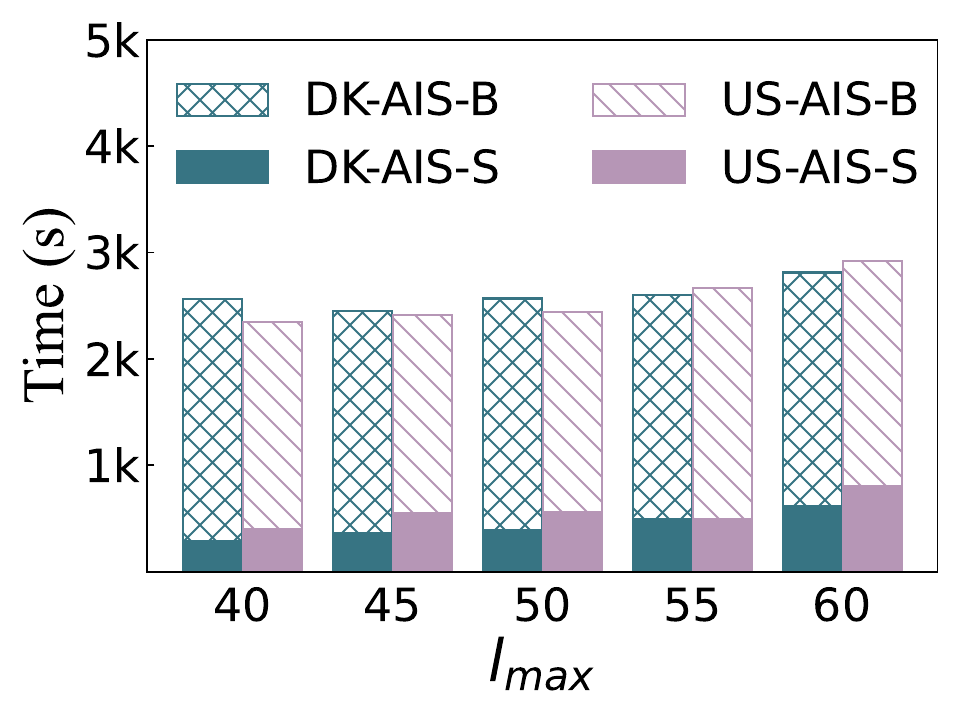}
            \label{fig:index_ln_time}
        \end{minipage}
    }
    \subfigure[Size vs. $l_\textit{max}$]{
        \begin{minipage}[t]{0.45\columnwidth}
            \centering
            \includegraphics[width=\columnwidth]{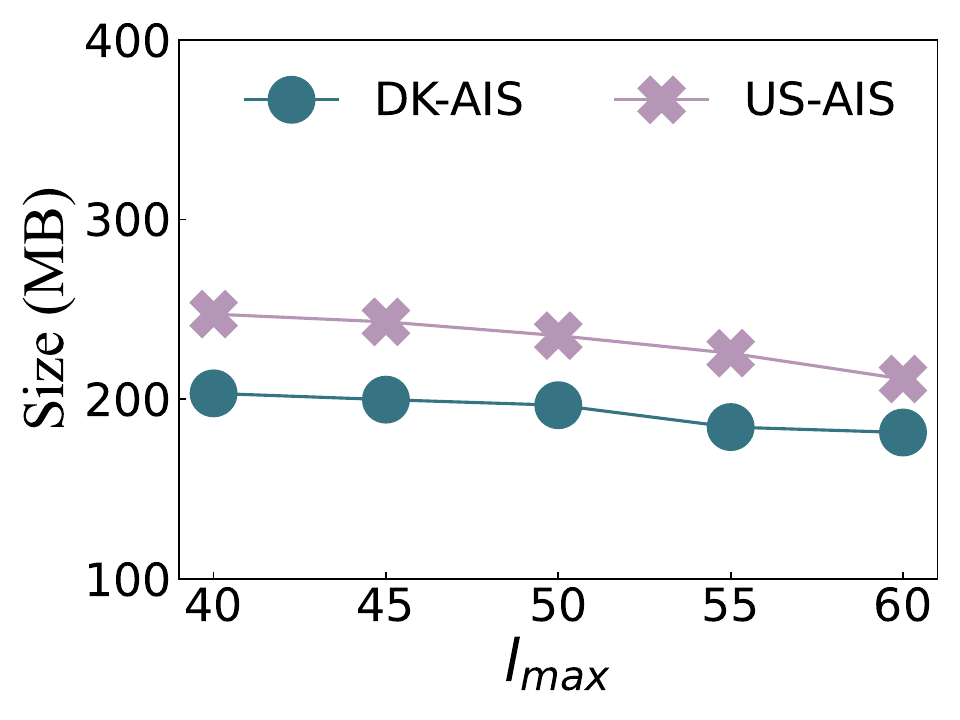}
            \label{fig:index_ln_size}
        \end{minipage}
    }
    \centering
    \caption{Effect of $l_\textit{max}$ ($l_\text{min} = 30$).}
    \label{fig:index_ln}
\end{figure}
We evaluate the efficiency of \texttt{SVTI} construction when varying different segmentation parameter settings.
First, we vary the minimum segment length $l_\text{min}$, keeping the maximum segment length $l_\textit{max}$ at $l_\text{min} + 20$. As $l_\text{min}$ increases, the time consumption in the segmentation phase, i.e., \texttt{DK-AIS-S} and \texttt{US-AIS-S}, shows a slight increase for across both datasets. This occurs because larger values of $l_\text{min}$ require more iterations in the segmentation process.
However, the total construction time and index size decrease as $l_\text{min}$ increases. This is due to the generation of fewer segments, which subsequently reduces both the time required for construction and the overall index size.

We also vary the maximum segment length $l_\textit{max}$, while fixing the minimum segment length $l_\text{min}$ at 30. The results are shown in Fig.~\ref{fig:index_ln}. An increase in $l_\textit{max}$ results in longer segmentation times for both datasets, as the process requires more iterations with larger $l_\textit{max}$ values.
Additionally, the number of generated segments decreases slightly as $l_\textit{max}$ grows, leading to a modest reduction in both index size and the construction time during the building phase, i.e., \texttt{DK-AIS-B} and \texttt{US-AIS-B}.

\vfill
\end{document}